\documentclass[twocolumn,aps,PRL]{revtex4}
\usepackage[latin9]{inputenc}
\usepackage{color}
\usepackage{amsmath}
\usepackage{amssymb}
\usepackage{graphicx}

\makeatletter

\providecommand{\tabularnewline}{\\}

\@ifundefined{textcolor}{}
{%
 \definecolor{BLACK}{gray}{0}
 \definecolor{WHITE}{gray}{1}
 \definecolor{RED}{rgb}{1,0,0}
 \definecolor{GREEN}{rgb}{0,1,0}
 \definecolor{BLUE}{rgb}{0,0,1}
 \definecolor{CYAN}{cmyk}{1,0,0,0}
 \definecolor{MAGENTA}{cmyk}{0,1,0,0}
 \definecolor{YELLOW}{cmyk}{0,0,1,0}
}

\def\NOT(#1,#2){\OneQubitGate(#1,#2){$X$}}

\makeatother

\begin{document}
\title{Efficient quantum gates for individual nuclear spin qubits by indirect
control }
\author{Swathi S. Hegde, Jingfu Zhang, and Dieter Suter\\
 Fakultät Physik, Technische Universität Dortmund,\\
 D-44221 Dortmund, Germany }
\begin{abstract}
Hybrid quantum registers, such as electron-nuclear spin systems, have
emerged as promising hardware for implementing quantum information
and computing protocols in scalable systems. Nevertheless, the coherent
control of such systems still faces challenges. Particularly, the
lower gyromagnetic ratios of the nuclear spins cause them to respond
slowly to control fields, resulting in gate times that are generally
longer than the coherence time of the electron. Here, we demonstrate
a scheme for circumventing this problem by indirect control: We apply
a small number of short pulses only to the electron and let the full
system undergo free evolution under the hyperfine coupling between
the pulses. Using this scheme, we realize robust quantum gates in
an electron-nuclear spin system, including a Hadamard gate on the
nuclear spin and a controlled-NOT gate with the nuclear spin as the
target qubit. The durations of these gates are shorter than the electron
coherence time, and thus additional operations to extend the system
coherence time are not needed. Our demonstration serves as a proof
of concept for achieving efficient coherent control of electron-nuclear
spin systems, such as NV centers in diamond. Our scheme is still applicable
when the nuclear spins are only weakly coupled to the electron. 
\end{abstract}
\maketitle
Spin-based quantum registers have come up as a feasible architecture
for implementing quantum computing \cite{nielsen,stolze}. Among them
are the hybrid systems consisting of electron and nuclear spins such
as Nitrogen Vacancy (NV) centers in diamond \cite{gaebel2006room,neumann2008multipartite,wrachtrup2001quantum,suter2017single,childress2006coherent,fuchs2009gigahertz,balasubramanian2009ultralong,maurer2012room,herbschleb2019ultra,bradley2019ten,gali2008ab}.
Specific properties of their subsystems are the distinct gyromagnetic
ratios, which result, e.g. in the requirement that the frequencies
of the control fields applied to electronic and nuclear spins lie
in the microwave (MW) and radiofrequency (RF) regimes respectively.
The fast gate operation times on the electrons (order of ns) and the
long coherence times of the nuclear spins (order of ms) serve as efficient
control and memory channels. However, the lower gyromagnetic ratios
of the nuclear spins result in longer nuclear spin gate operation
times (a few tens of $\mu$s), which can exceed the electron coherence
times ($\approx1-25$ $\mu$s) at room temperature, thus posing a
major challenge for coherent control of electron-nuclear spin systems.
Techniques like dynamical decoupling (DD) can partly alleviate this
issue by extending the coherence times of the electron \cite{meiboom1958modified,uhrig2007keeping,zhang2015experimental,van2012decoherence,RevModPhys.88.041001,zhang2014protected},
but the additional DD pulses increase the control cost.

Previously, one- and two-qubit operations were demonstrated using
RF pulses on the nuclear spin that had strong hyperfine coupling of
$\approx130$ MHz \cite{jelezko2004observation,shim2013room,rao2016characterization}.
Such strong couplings enhance the nuclear spin Rabi frequency allowing
fast RF operations (order of ns) and hence direct control of nuclear
spins was feasible \cite{maze2008electron,shim2013room}. However,
scalable quantum computing requires coherent control of tens to hundreds
of qubits and the control of dipolar coupled nuclear spins gets challenging
with increasing distance from the electrons. To avoid these challenges,
indirect control (IC) of the nuclear spins has also been incorporated
\cite{khaneja2007switched,wang2017room,zhang2011coherent,hodges2008universal,cappellaro2009coherence,taminiau2012detection,aiello2015time}.
In this approach, the control fields are applied only on the electron,
combined with free evolution of the system under the hyperfine couplings.
However, most of the earlier works based on IC required a large number
of control operations, thereby increasing the control overhead \cite{taminiau2014universal,hodges2008universal}.

In this letter, we experimentally implement efficient quantum gates
in an NV center in diamond at room temperature, using IC with minimal
control cost of only 2-3 of short MW pulses and delays. Our approach
allows variable delays and pulse parameters. As such, it differs from
earlier work \cite{taminiau2014universal} that used many DD cycles
with fixed delays. We use this approach to demonstrate quantum gates
that are required for a universal set of gates: a Hadamard gate on
a nuclear spin, and a controlled-NOT (CNOT) gate with control on the
electron and target on the nuclear spin.

We consider a single NV center that consists of a spin-1 electron
coupled to a spin-1 $^{14}$N and a spin-$1/2$ $^{13}$C [see supplementary
material \cite{note6}]. We perform the operations on the electron and $^{13}$C
by focussing on a subspace of the system where the $^{14}$N is in
the $m_{N}=1$ state. We then can write the secular part of the electron-$^{13}$C
Hamiltonian in the lab frame as ${\cal {H}}/(2\pi)=D(S_{z}^{2}\otimes E_{2})-(\nu_{e}-A_{N})(S_{z}\otimes E_{2})-\nu_{C}(E_{3}\otimes I_{z})+A_{zz}(S_{z}\otimes I_{z})+A_{zx}(S_{z}\otimes I_{x}),$
where $S_{z}$ and $I_{z/x}$ are the spin operators for electron
and $^{13}$C respectively, $E_{n}$ is an $n\times n$ identity matrix,
$D=2.87$ GHz is the zero field splitting, $\nu_{e}=-414$ MHz and
$\nu_{C}=0.158$ MHz are the Larmor frequencies of the electron and
$^{13}$C in a $14.8$ mT field, $A_{N}=-2.16$ MHz is the hyperfine
coupling with $^{14}$N and $A_{zz}=-0.152$ MHz and $A_{zx}=0.110$
MHz are the hyperfine couplings with $^{13}$C. The eigenstates of
${\cal {H}}$ are $|0\uparrow\rangle,|0\downarrow\rangle,|-1\varphi_{-}\rangle,|-1\psi_{-}\rangle,|1\varphi_{+}\rangle,|1\psi_{+}\rangle$,
where $\{|0\rangle,|\pm1\rangle\}$ are the eigenstates of $S_{z}$,
and

\begin{eqnarray}
|\varphi_{\pm}\rangle & = & \cos(\kappa_{\pm}/2)|\uparrow\rangle+\sin(\kappa_{\pm}/2)|\downarrow\rangle\nonumber \\
|\psi_{\pm}\rangle & = & -\sin(\kappa_{\pm}/2)|\uparrow\rangle+\cos(\kappa_{\pm}/2)|\downarrow\rangle.\label{phi-2}
\end{eqnarray}
Here $\{|\uparrow\rangle$,$|\downarrow\rangle\}$ are the eigenstates
of $I_{z}$, and $\kappa_{\pm}=\arctan[A_{zx}/(A_{zz}\mp\nu_{C})]$
is the angle between the quantization axis of the $^{13}$C and the
NV axis.

We implement the quantum gates $U_{T}$ in the $m_{S}=\{0,-1\}$ and
$m_{N}=1$ manifold and refer to it as the system subspace.
 This choice of subspace is realized by using MW pulses with a Rabi
frequency of $\approx0.5$ MHz ($\ll A_{N})$, which covers all ESR
transitions in the system subspace but leaves states untouched where
the $^{14}$N is in a different state. For the system subspace, the
Hamiltonian is ${\cal H}_{s}/(2\pi)=|0\rangle\langle0|\otimes{\cal H}_{0}+|-1\rangle\langle-1|\otimes{\cal H}_{-1}$,
where ${\cal H}_{0}=-\nu_{C}I_{z}$ and ${\cal H}_{-1}=-(\nu_{C}+A_{zz})I_{z}-A_{zx}I_{x}$
are $^{13}$C spin Hamiltonians when the electron is in $|0\rangle$
or $|-1\rangle$ respectively.

We implement two examples of $U_{T}$: 
\begin{eqnarray}
U_{H} & = & E_{2}\otimes\begin{bmatrix}1 & 1\\
1 & -1
\end{bmatrix}/\sqrt{2}\nonumber \\
U_{CNOT} & = & |0\rangle\langle0|\otimes E_{2}+|-1\rangle\langle-1|\otimes e^{-i\pi I_{x}}.\label{ut}
\end{eqnarray}

The first is a Hadamard gate while the second is a CNOT gate, both
targeting $^{13}$C, in a basis defined in Ref. \cite{note5}. To
check the implementation of $U_{T}$, we initialize the system into
a pure state, apply $U_{T}$ and then perform a partial tomography
of the final state by recording free precession signals (FIDs).

\begin{table}[b]
\centering %
\begin{tabular}{|c|c|c|c|c|c|c|c|c|c|c|}
\hline 
 & $\tau_{1}$  & $\tau_{2}$  & $\tau_{3}$  & $\tau_{4}$  & $t_{1}$  & $t_{2}$  & $t_{3}$  & $\phi_{1}$  & $\phi_{2}$  & $\phi_{3}$ \tabularnewline
\hline 
$U_{H}$  & $0.74$  & $0.22$  & $0.43$  & $0.89$  & $0.23$  & $1.26$  & $1.50$  & $3\pi/2$  & $3\pi/2$  & $\pi/2$ \tabularnewline
\hline 
$U_{CNOT}$  & $3.78$  & $2.11$  & $2.15$  & $0.63$  & $1.88$  & $3.96$  & $1.90$  & $0$  & $\pi/5$  & $\pi/2$ \tabularnewline
\hline 
\end{tabular}\caption{MW pulse sequence parameters for $U_{H}$ and $U_{CNOT}$. The time
durations and phases are in units of $\mu$s and radians respectively.}
\label{tab1} 
\end{table}

For practical applications, it is useful to allow additional degrees
of freedom, such as variable pulse rotation angles and finite pulse
durations. These degrees of freedom allow us to compensate experimental
errors via numerical optimization of the pulse sequence parameters.
As shown in Fig. \ref{bloch}, we consider a pulse sequence consisting
of delays $\tau_{i}$ and MW pulses with durations $t_{i}$ and phases
$\phi_{i}$ where $i=1\cdots n$, $n$ is the number of pulses. We
fix the frequency of the pulses to be resonant with the ESR transition
$0\leftrightarrow-1$ and the Rabi frequency $\omega_{1}/2\pi$ to
$0.5$ MHz. During $\tau_{i}$, the system freely evolves under ${\cal {H}}_{s}$
such that $U_{i}^{f}=e^{-i{\cal {H}}_{s}\tau_{i}}$. The control Hamiltonians
during the MW pulse segments are ${\cal {H}}_{i}^{MW}=\omega_{1}[\cos\phi_{i}(s_{x}\otimes E_{2})+\sin\phi_{i}(s_{y}\otimes E_{2})]+{\cal {H}}_{s}$,
where $s_{x/y}$ denote the spin-1/2 operators for the electron, and
the corresponding operators are $U_{i}^{MW}=e^{-i{\cal {H}}_{i}^{MW}t_{i}}$.
The total propagator $U$ is the time ordered product of $U_{i}^{f}$
and $U_{i}^{MW}$. The overlap between $U$ and $U_{T}$ is defined
by the fidelity $F=|\mathrm{Tr}(U^{\dagger}U_{T})|/4$. We maximize
$F$ numerically, using a MATLAB$^{\tiny{\textregistered}}$ subroutine
implementing a genetic algorithm \cite{Mitchell:1998:IGA:522098}.
The solution returns the pulse sequence parameters $t_{i}$, $\tau_{i}$
and $\phi_{i}$. The sequences were made robust against fluctuations
of the MW pulse amplitude by optimizing $F$ over a range $\omega_{1}/(2\pi)=[0.48,0.52]$
MHz. Table \ref{tab1} summarizes the optimized pulse parameters for
$U_{H}$ and $U_{CNOT}$, and the average gate fidelities are $>96\%$
and $>97\%$ respectively. The resulting trajectories of the electron
and $^{13}$C on the Bloch-sphere is shown in the SM \cite{note6}.

Our experiments started with an initial laser pulse with a wavelength
of $532$ nm, a duration of $5$ $\mu$s, and a power of $\approx0.5$
mW which initialized the electron to $|0\rangle$ but left the $^{13}$C
in a mixed state. To initialize $^{13}$C to $|\uparrow\rangle$,
we resorted to the IC method \cite{zhang2018efficient,zhangcnot,note6}. Starting from $\psi_{0}=|0\uparrow\rangle$, we implemented
the circuits shown in Figs. (\ref{had}, \ref{ps}). 
 Depending on the experiment, we either observed the electron or the
$^{13}$C state via FID measurements. The readout process consisted
of another laser pulse with the same wavelength and $400$ ns duration
and was used to measure the population of $m_{S}=0$.

\begin{figure}[t]
\centering \includegraphics[width=6.1cm]{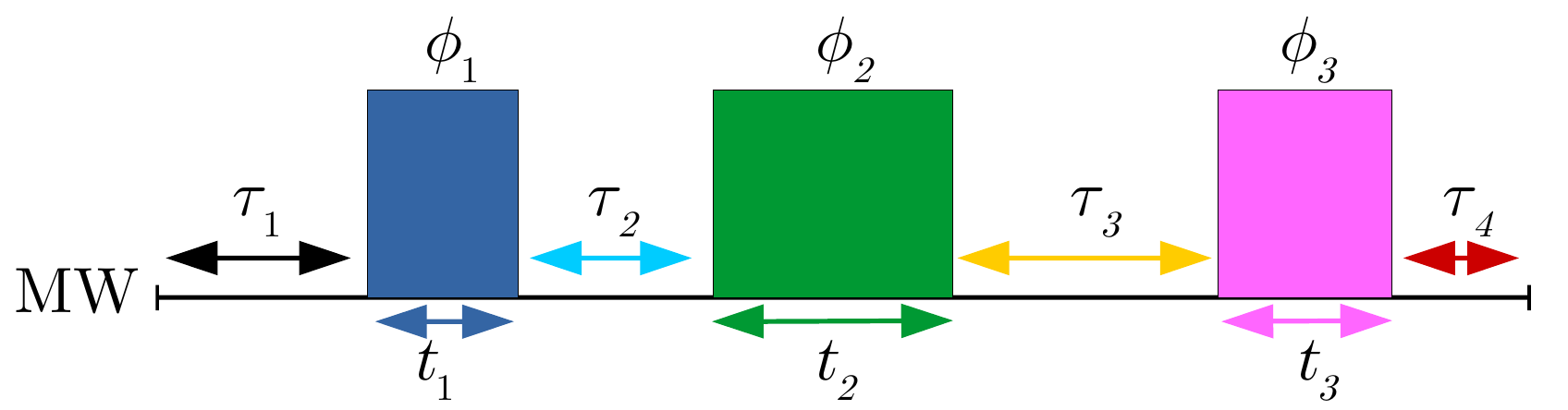} \caption{MW pulse sequence to realize $U_{T}$ by IC, at a fixed $\omega_{1}$.
The delays $\tau_{i}$, MW pulse durations $t_{i}$ and phases $\phi_{i}$
are the free variables to be optimized.}
\label{bloch} 
\end{figure}

\begin{figure}[b]
\centering \includegraphics[width=3.6cm]{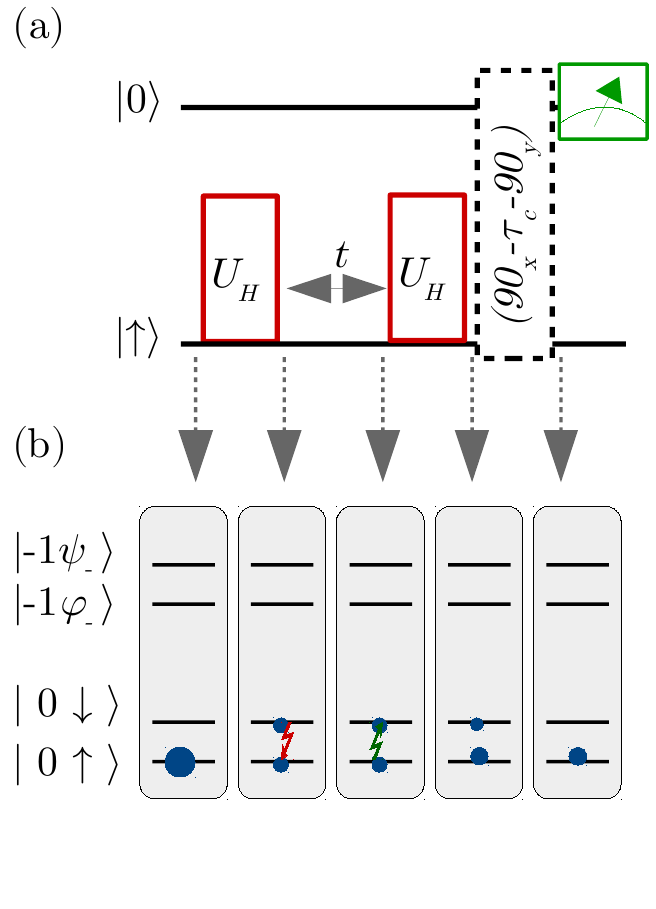} \includegraphics[width=4.5cm]{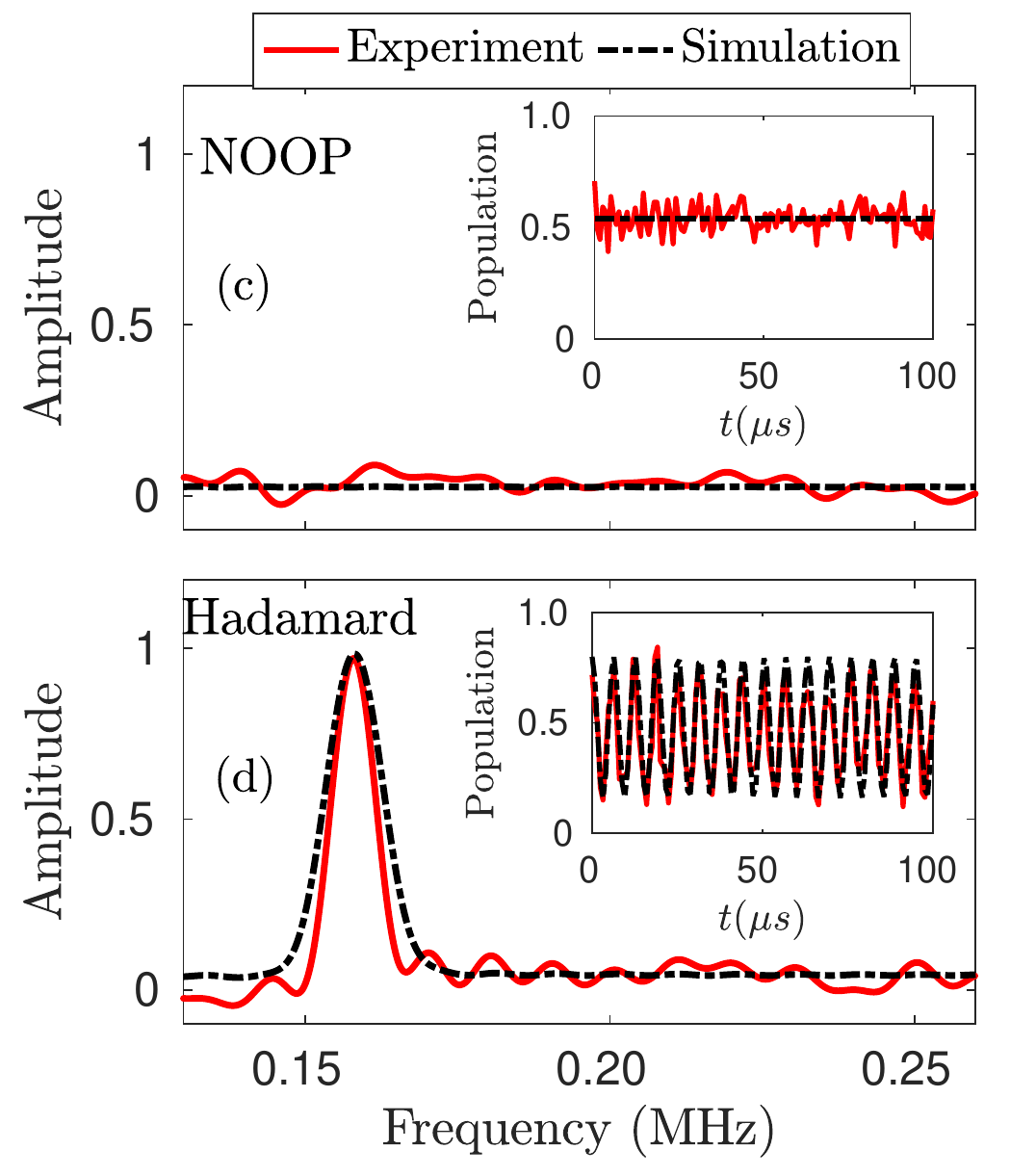}
\caption{(a) Quantum circuit to test $U_{H}$. The MW pulse sequence parameters
for $U_{H}$ are given in Table. \ref{tab1}. The clean-up operation
 is represented by the dotted box. (b) Populations (solid circles)
and coherences (zig-zag arrows) at each stage of the pulse sequence
in (a). (c, d) $^{13}$C spin spectra obtained by the pulse sequence
in (a). (c) Without the first $U_{H}$. (d) With both $U_{H}$. Inset:
Final population of $|0\uparrow\rangle$ as a function of $t$. }
\label{had} 
\end{figure}

Fig. \ref{had}(a) shows the pulse sequence for implementing and detecting
the effect of $U_{H}$. The first $U_{H}$ generates $|0\rangle\otimes(|\uparrow\rangle+|\downarrow\rangle)/\sqrt{2}$.
The $^{13}$C coherence is then allowed to evolve for a variable time
$t$ after which we apply another $U_{H}$ to convert one component
of the coherence to population. Lastly, a clean-up operation, with
MW pulse sequence $(90_{x}-\tau_{c}-90_{y})$, where $90_{x/y}$ are
pulses with rotation angle $90^{\circ}$ about the $x/y$-axis applied
to the $m_{S}=0\leftrightarrow1$ transition with 0.5 MHz Rabi frequency
and $\tau_{c}=1/(2|A_{zz}|)$ is the delay, represented by the dotted
box transfers the population from $|0\downarrow\rangle$ to $|1\downarrow\rangle$.
The final read-out operation thus detects only the population of $|0\uparrow\rangle$,
which depends on $t$ as $[1+\cos(2\pi\nu_{C}t)]/2$. In the frequency
domain, this corresponds to a peak at $\nu_{C}$.

Using the pulse sequence in Fig. \ref{had}(a), we performed two experiments
to compare the effect of $U_{H}$: (1) without the first $U_{H}$
(i.e, no operation, also known as NOOP) and (2) with both $U_{H}$.
In the case of NOOP, the system was in $\psi_{0}$ during the free
evolution period. Since $\psi_{0}$ does not contain $^{13}$C coherence
the resulting frequency domain signal does not contain a resonance
at $\nu_{C}$, as shown in Fig. \ref{had}(c). With both $U_{H}$
present, we observe in Fig. \ref{had}(d) a resonance peak at $\nu_{C}$
as expected. We numerically simulated the pulse sequence in Fig. \ref{had}(a)
without and with the first $U_{H}$, and then calculated the final
populations of $|0\uparrow\rangle$ as a function of $t$. To match
the theoretical signal with the experimental one, we had to scale
it by a factor $0.9$ for NOOP and $0.8$ for $U_{H}$ (i.e, with
two $U_{H}$), and estimated the infidelity 
 of the experimental $U_{H}$ as $\approx10\%$.

\begin{figure}[t]
\centering\includegraphics[width=8.5cm]{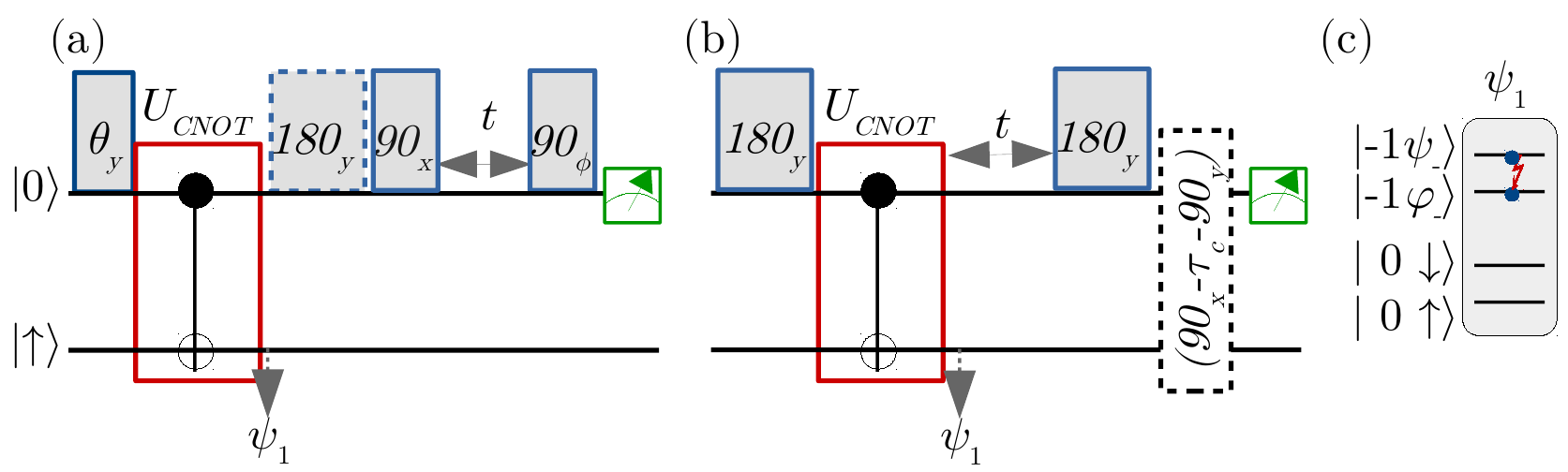} \caption{Quantum circuits to test $U_{CNOT}$. The MW pulse sequence parameters
for $U_{CNOT}$ indicated by red empty boxes are given in Table. \ref{tab1}.
$\theta_{x/y/\phi}$ denote operations with rotation angles $\theta$
about the $x/y/\phi$ axes that are resonant with the transition $0\leftrightarrow-1$
and with Rabi frequencies of $8$ MHz. (a) Pulse sequence to demonstrate
the effect of $U_{CNOT}$ on different input states via electron spin
detection. $\phi$ is the detuning phase. In the presence (absence)
of the $180_{y}$ operation indicated by the dashed box, the FID measurement
is used to determine the population of the $m_{S}=-1$ ($m_{S}=0$)
after $U_{CNOT}$. 
 (b) Pulse sequence to demonstrate the effect of $U_{CNOT}$ via $^{13}$C
spin detection. (c) Pictorial representation of state $\psi_{1}$. }
\label{ps} 
\end{figure}

\begin{figure}[t]
\centering\includegraphics[width=8.5cm]{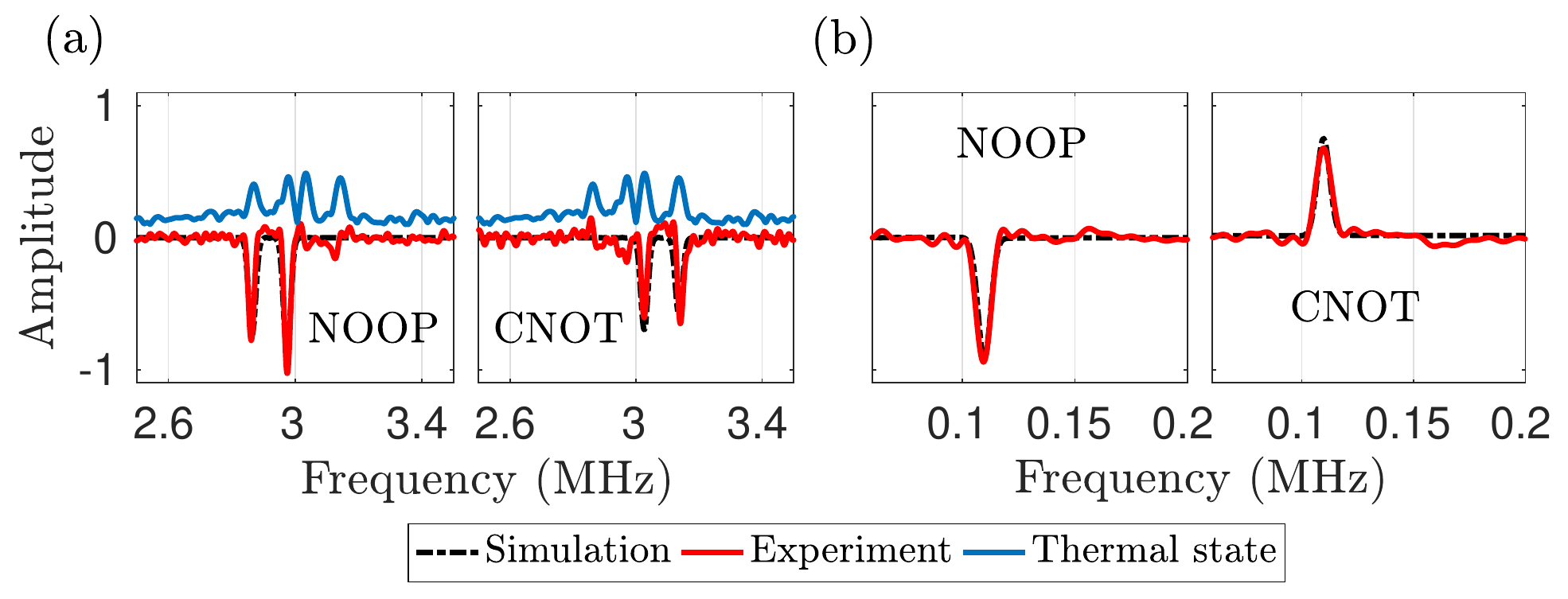} \caption{(a) Electron spin spectra for the pulse sequence corresponding to
Fig. \ref{ps}(a) without and with $U_{CNOT}$where $\theta_{y}=\pi$.
The thermal state spectra on top are shifted vertically for reference.
The electron spin spectra are centered around the detuning frequency
3 MHz. (b) $^{13}$C spin spectra obtained by the pulse sequence shown
in Fig. \ref{ps}(b) without and with $U_{CNOT}$. The peaks appear
at $\nu_{-}$= 0.11 MHz.}
\label{cnot_e} 
\end{figure}

The schemes to demonstrate $U_{CNOT}$\textcolor{red}{{} }are shown
in Fig. \ref{ps}. Using the pulse sequence in Fig. \ref{ps}(a),
we demonstrated the effect of $U_{CNOT}$ in $m_{S}=-1$ by measuring
electron spin spectra. Choosing for the flip-angle $\theta$ of the
initial $\theta_{y}$ operation \cite{note1,cavanagh1995protein} a value of $\pi$, we
exchanged the populations of the $|0\uparrow\rangle$ $\leftrightarrow|-1\uparrow\rangle\approx|-1\rangle\otimes(|\phi_{-}\rangle-|\psi_{-}\rangle)/\sqrt{2}$
according to Eq. (\ref{phi-2}). The subsequent $U_{CNOT}$ transformed
$|-1\uparrow\rangle$ to $-i|-1\downarrow\rangle\approx-i|-1\rangle\otimes(|\phi_{-}\rangle+|\psi_{-}\rangle)/\sqrt{2}$,
since by definition of Eq. (\ref{ut}), $U_{CNOT}$ flips the $^{13}$C
state when the electron is in $|-1\rangle$. To measure the state
after $U_{CNOT}$, we transferred the population of $|-1\downarrow\rangle$
to $|0\downarrow\rangle$ using a hard $180_{y}$ operation. The readout
process, which measures the population of $m_{S}=0$, can then be
used to determine the population left in $|-1\downarrow\rangle$ by
$U_{CNOT}$. The sequence $(90_{x}-t-90_{\phi})$ in Fig. \ref{ps}(a)
implements the electron spin FID measurement, where the $90_{x}$
pulse creates electron coherence and the $90_{\phi}$ pulse converts
one component of the evolved coherence to population \cite{suter2017single,zhangcnot}.
Here we incremented the phase $\phi(t)=-2\pi\nu_{d}t$ linearly with
$t$, using a detuning frequency $\nu_{d}$ of 3 MHz. We then measured
the population of $m_{S}=0$ with the readout laser pulse as a function
of $t$ and its Fourier transform gives the frequency domain signal.
Thus, as seen in the electron spin spectra in Fig. \ref{cnot_e}(a),
the change of nuclear spin state resulted in a different frequency
of the ESR lines in the case of $U_{CNOT}$ as compared to NOOP.

Since $U_{CNOT}$ targets the $^{13}$C, we also observed its effects
on the $^{13}$C by measuring the nuclear spin spectra using the pulse
sequence in Fig. \ref{ps}(b). The initial $180_{y}$ operation transforms
$|0\uparrow\rangle$ to $|-1\uparrow\rangle\approx|-1\rangle\otimes(|\varphi_{-}\rangle-|\psi_{-}\rangle)/\sqrt{2}$.
After implementing $U_{CNOT}$, we allowed the $^{13}$C coherence
between states $|\varphi_{-}\rangle$ and $|\psi_{-}\rangle$ to evolve
for a variable time $t$, as shown in Fig. \ref{ps}(c), and then
applied another $180_{y}$ operation to the electron to bring the
evolved state from $m_{S}=-1$ to $m_{S}=0$. The subsequent clean-up
operation removed the population of $|0\downarrow\rangle$ and allowed
us to measure the remaining population of $|0\uparrow\rangle$ with
the readout laser pulse. The experimental $^{13}$C spectra without
and with $U_{CNOT}$ are shown in Fig. \ref{cnot_e}(b). The resonance
frequency of the peak at $0.11$ MHz agree with the expected resonance
frequency $\nu_{-}$ of the $^{13}$C for $m_{S}=-1$. Comparing with
NOOP, the inverted amplitude shows that $U_{CNOT}$ flipped the $^{13}$C
states in $m_{S}=-1$. In Figs. \ref{cnot_e}(a, b), we show the matching
simulations, calculated for ideal pulses, scaled by a factor $0.8$.

\begin{figure}[b]
\centering\includegraphics[width=8.7cm]{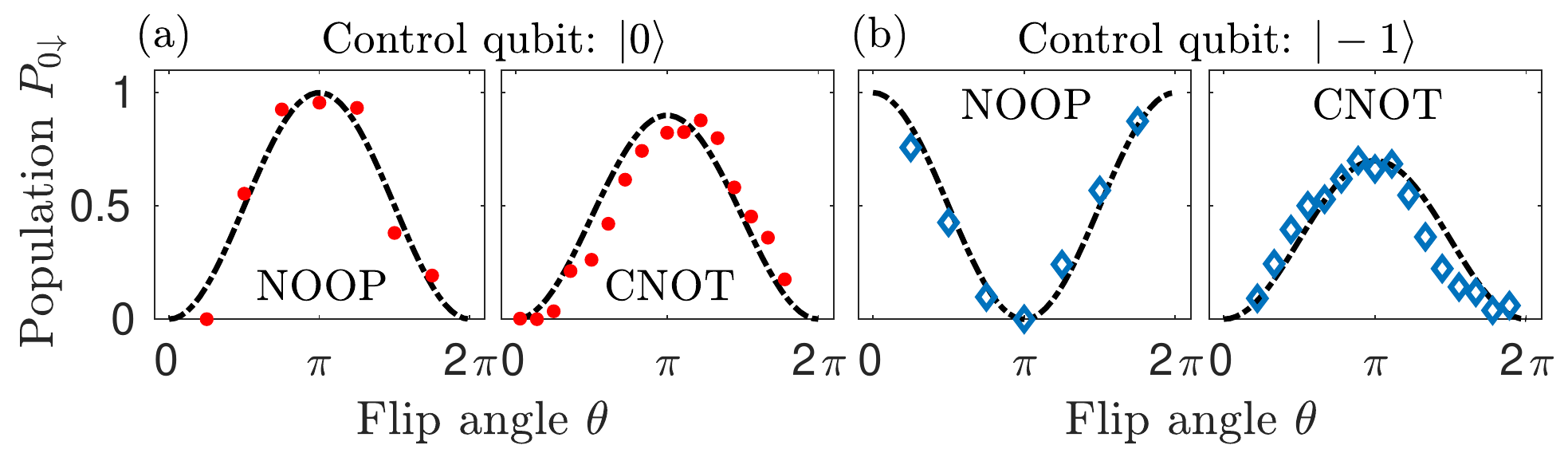} \caption{$P_{0\downarrow}$ as a function of $\theta$ corresponding to the
pulse sequences shown in Fig. \ref{ps}(a). The diamonds and solid
circles are the experimental data, and the dashed lines are the matching
simulations.}
\label{cnot_e-2} 
\end{figure}

As an additional test of the sequence for different input states,
we first applied a selective rotation, when $m_{N}=1$ \cite{note2},
of $\psi_{0}$ by an angle $\theta_{y}$ to generate the superposition
state $\psi_{\theta}=\cos(\theta/2)|0\uparrow\rangle+\sin(\theta/2)|-1\uparrow\rangle$.
As shown in Fig. \ref{ps}(a), we then applied either a NOOP or $U_{CNOT}$.
The latter transforms $\psi_{\theta}$ to $\cos(\theta/2)|0\uparrow\rangle-i\sin(\theta/2)|-1\downarrow\rangle$,
which is entangled for $\theta\ne n\pi$ with integer $n$. \textcolor{black}{Ideally,
the amplitude of the resonance line for the transition $|0\downarrow\rangle$
$\leftrightarrow|1\downarrow\rangle$ \cite{note3} is proportional
to the population $P_{0\downarrow}$. We thus determined $P_{0\downarrow}$}
and the results, which are shown in Fig. \ref{cnot_e-2}, demonstrate
the effect of $U_{CNOT}$ for the 2 cases where the control qubit
is $|0\rangle$ or $|-1\rangle$. Fig. \ref{cnot_e-2}(a) shows $P_{0\downarrow}$
after applying NOOP or $U_{CNOT}$ to $\psi_{\theta}$, as a function
of $\theta$ in the absence of the $180_{y}$ operation indicated
by the dotted box in Fig. \ref{ps}(a). This pulse sequence allows
us to measure the effect of $U_{CNOT}$ when the electron spin is
$|0\rangle$. The curves for both cases are similar since $U_{CNOT}$
does not change the $^{13}$C state when the electron spin is $|0\rangle$.
In Fig. \ref{cnot_e-2}(b) we show the effect of $U_{CNOT}$ when
the electron spin is $|-1\rangle$. To read out the population of
$|-1\downarrow\rangle$, we first applied a $180_{y}$ operation,
as shown in Fig. \ref{ps}(a) and then measured the electron spin
FID in $m_{S}=\{0,1\}$. In this case, the $P_{0\downarrow}$ vs $\theta$
curve flipped for $U_{CNOT}$ compared to NOOP, indicating the change
of the $^{13}$C state when the electron is in $|-1\rangle$. By fitting
the experimental $P_{0\downarrow}$ with the corresponding theoretical
populations for various $\theta$ as shown in Fig. \ref{cnot_e-2},
we estimated the experimental infidelity due to $U_{CNOT}$ as $20\%$
 \cite{note6}.

\textit{{Discussion.---}}{{} Our experiments convincingly show
that the IC scheme is a very effective approach to implement operations
in systems consisting of 3 types of qubits. The advantages of this
approach will become even more important as the number of qubits increases.
While a full implementation of the approach in large quantum registers
is beyond the scope of this paper, we have tested the basic scheme
through numerical simulations of gates in multiqubit systems with
up to six qubits. The simulations show that the procedure scales relatively
favorably with the size of the system \cite{note6}. For the 6-qubit
system our method to control individual $^{13}$C spins was efficient
as it required 3-4 MW pulses and the total duration was $<30\,\mu$s.
The theory \cite{khaneja2007switched, lowenthal1971uniform} regarding the bounds for the
control overhead and the condition to retain efficiency for larger
spin systems is explained in \cite{note6}.

\textit{Conclusion.---} We experimentally demonstrated full coherent
control i.e, state initialization, gate implementation and detection
of the electron-nuclear spin system in the NV center of diamond using
the methods of IC. We specifically chose a center with a small hyperfine
coupling, some three orders of magnitude weaker than that of the nearest
neighbor $^{13}$C spins. The distance between the electron and $^{13}$C
is $\approx0.89$ nm \cite{note6}. These remote spins are much more abundant than
the nearest neighbors and their relaxation times much longer. However,
since their coupling to RF fields is also much weaker, direct RF excitation
does not lead to efficient control operations. The IC techniques that
we have demonstrated allow much faster controls and therefore overall
higher fidelity - an essential prerequisite for scalable quantum systems.
Specifically, we have implemented a Hadamard gate on $^{13}$C and
a CNOT gate, where the electron is the control qubit and $^{13}$C
the target qubit, using only a small number of MW pulses and delays.
The above gate operations targeted the subspace $m_{S}=\{0,-1\}$
and $m_{N}=1$. If we consider the control state of the $^{14}$N,
i.e $m_{N}=1$, in the whole space with $m_{N}=\{{0,-1,1}\},$ then
our $U_{CNOT}$ is a Toffolli gate in $12$ dimensions. Since the
total duration of the pulse sequence was well within the electron
coherence time ($T_{2}^{*}\approx20\mu$s), additional coherence preserving
control operations were not required. However, for complex algorithms
consisting of many gates, it may be necessary to include DD. While
we have implemented this scheme in the diamond NV center at room temperature
in a small external magnetic field, it remains applicable over a much
wider parameter range and can clearly be adapted to other quantum
systems, thus opening the ways for many different implementations
of advanced quantum algorithms using indirect control schemes.

\textit{Acknowledgments.---} This work was supported by the DFG through
grants SU 192/34-1 and SU 192/31-1 and by the European Union's Horizon 2020 research and innovation programme 
under grant agreement No 828946. The publication reflects the opinion
of the authors; the agency and the commission may not be held responsible
for the information contained in it. SH thanks Dr T S Mahesh for fruitful
discussions on genetic algorithms.

 \bibliographystyle{apsrev}
\bibliography{refcnot3}

\begin{thebibliography}{44}
\bibitem[{not({\natexlab{b}})}]{note5}
\bibinfo{note}{These operations are written in the computational basis states
  $\{|0\uparrow\rangle,|0\downarrow\rangle,|-1\uparrow\rangle,|-1\downarrow\rangle\}$
  which is related to the energy eigenbasis by a transformation matrix
  $V=|0\rangle\langle0|\otimes E_{2}+|-1\rangle\langle-1|\otimes
  e^{-i\kappa_{-}I_{y}}$.}
\bibitem[{not({\natexlab{c}})}]{note1}
\bibinfo{note}{Here the hard $\theta_y$ operation refers to the operation in
  subspace $m_S =\{0,-1\}$ and $m_N =\{0,-1,1\}$. The corresponding Rabi
  frequency is 8MHz.}
\bibitem[{not({\natexlab{d}})}]{note2}
\bibinfo{note}{If $\theta_y$ operation is a hard pulse as before, then for any
  value of $\theta$ not equal to integral multiple of $\pi$, this pulse creates
  electron spin coherence in $m_N=\{0,-1,1\}$ subspaces that evolve during the
  gate operations implemented in the system subspace. For simplicity, we here
  chose $\theta_y$ operation as selective pulse subjected to $m_N=1$ with Rabi
  frequency 0.5 MHz as we vary $\theta$ value.}
\bibitem[{not({\natexlab{e}})}]{note3}
\bibinfo{note}{Unlike the 4 ESR peaks in $m_S=\{0,-1\}$ subspace, the ESR
  spectra in $m_S=\{0,1\}$ subspace has only two observable resonance peaks,
  where one peak corresponds to total population of state $|\uparrow\rangle$
  and the other peak corresponds to total population of state
  $|\downarrow\rangle$ [see Ref. 31].}
\end{thebibliography}

\pagebreak
\widetext
\begin{center}
\textbf{\large Supplemental Material for ``Efficient quantum gates for individual nuclear spin qubits by indirect
control"}\\
Swathi S. Hegde, Jingfu Zhang, and Dieter Suter\\
 Fakultät Physik, Technische Universität Dortmund,\\
 D-44221 Dortmund, Germany
\end{center}

\makeatletter
\section*{1. NV center system and  Bloch Sphere representation of the evolution}

\begin{figure}[h]
\centering %
\begin{tabular}{c}
\includegraphics[width=12cm]{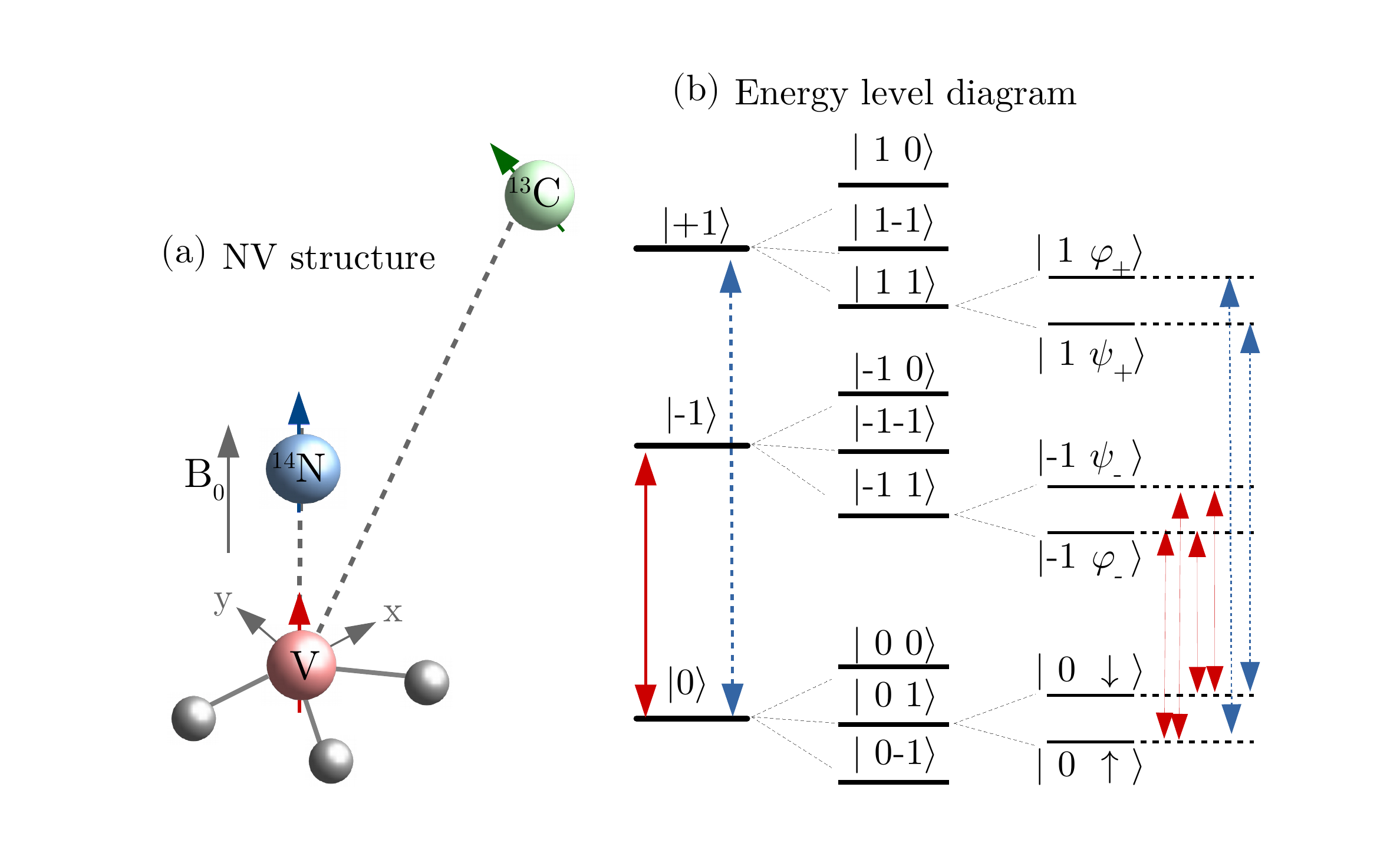} \tabularnewline
\end{tabular}\caption{(a) NV center coupled to $^{14}$N and $^{13}$C spins. The magnetic
field $B_{0}$ is aligned along the NV axis in the z direction. (b)
The energy level splittings. $\{|0\rangle,|\pm1\rangle\}$ correspond
to electron and $^{14}$N spins, $\{|\uparrow\rangle,|\downarrow\rangle,|\varphi_{\pm}\rangle,|\psi_{\pm}\rangle\}$
correspond to the $^{13}$C spin. The ESR transitions in the electron
spin subspace $m_{S}=$ $\{0,-1\}$, when $^{14}$N spin is in state
$m_{N}=1$, are shown by four arrows (red solid lines) in the right-hand
part. Similarly, the ESR transitions in the electron spin subspace
$m_{S}=$ $\{0,1\}$, when $^{14}$N spin is in state $m_{N}=1$,
are shown by two arrows (blue dotted lines) in the right-hand part.}
\label{nv} 
\end{figure}

The experiments were carried out on a diamond sample with $^{12}$C
enrichment of $99.995\%$, at room temperature and at a field strength
of 14.8 mT. The $T_{2}^{*}$ of the electronic spin that we used in
this experiment was about $\approx20$ $\mu$s. Fig. \ref{nv}(a)
shows the structure of a single NV center coupled to $^{14}$N and
$^{13}$C nuclear spins. The Hamiltonian ${\cal {H}}$ of this system
 is discussed
in the main manuscript. Fig. \ref{nv}(b) shows the corresponding
energy level diagram. The external magnetic field of strength $B_{0}=14.8$
mT lifts the degeneracy of the electronic $|-1\rangle$ and $|+1\rangle$
states. Each of the spin-1 electronic states splits into $\{|0\rangle,|\pm1\rangle\}$
states of the $^{14}$N spin, which further split into the states
$\{|\uparrow\rangle,|\downarrow\rangle,|\varphi_{\pm}\rangle,|\psi_{\pm}\rangle\}$
of the $^{13}$C spin.

We chose a subspace where the electron spin was in $m_{S}=$ $\{0,-1\}$
and the $^{14}$N spin was in $m_{N}=1$ and referred to this subspace
as our system subspace in which we implemented our gate operations.
In the system subspace, there are 4 ESR transitions as indicated by
the red arrows in Fig. \ref{nv}(b), since the states $|\varphi_{-}\rangle$
and $|\psi_{-}\rangle$ are linear combinations of $|\uparrow\rangle$
and $|\downarrow\rangle$ states with $\kappa_{-}\approx86^{\circ}$
as described in Eqs. (1, 2) of the main manuscript.

In the subspace $m_{S}=$ $\{0,1\}$ when $m_{N}=1$, we observe that
$\kappa_{+}\approx10^{\circ}$. Eqs. (1, 2) of the main manuscript
indicate that $|\varphi_{+}\rangle\approx|\uparrow\rangle$ and $|\psi_{+}\rangle\approx|\downarrow\rangle$.
Therefore only 2 ESR transitions are observed for $m_{S}=$ $\{0,1\}$
and $m_{N}=1$, which correspond to the transitions $|0\uparrow\rangle\leftrightarrow|1\uparrow\rangle$
and \textbar$0\downarrow\rangle\leftrightarrow|1\downarrow\rangle${}
as indicated by the blue arrows in Fig. \ref{nv}(b). This subspace
was used to implement the clean-up operation.

Fig. \ref{bloch}(a) is our generic 3-pulse sequence for implementing
$U_{H}$ and $U_{CNOT}$. Fig. \ref{bloch}(b,c) shows the resulting
trajectories of the electron and $^{13}$C on the Bloch-sphere.

\begin{figure}[t]
\centering \includegraphics[width=6.1cm]{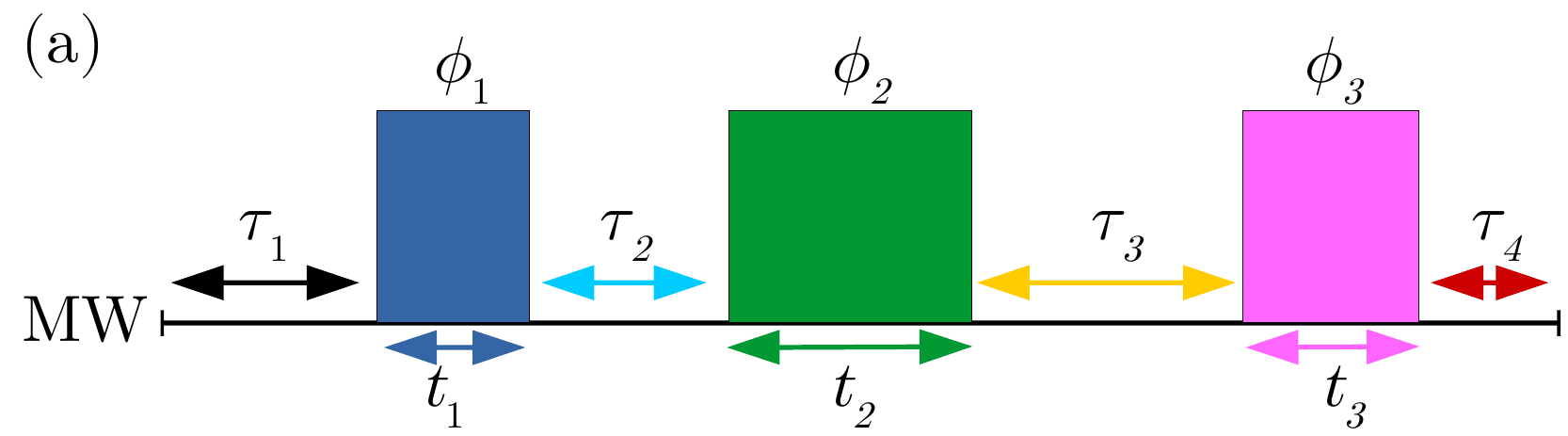}\\
 \includegraphics[width=8cm]{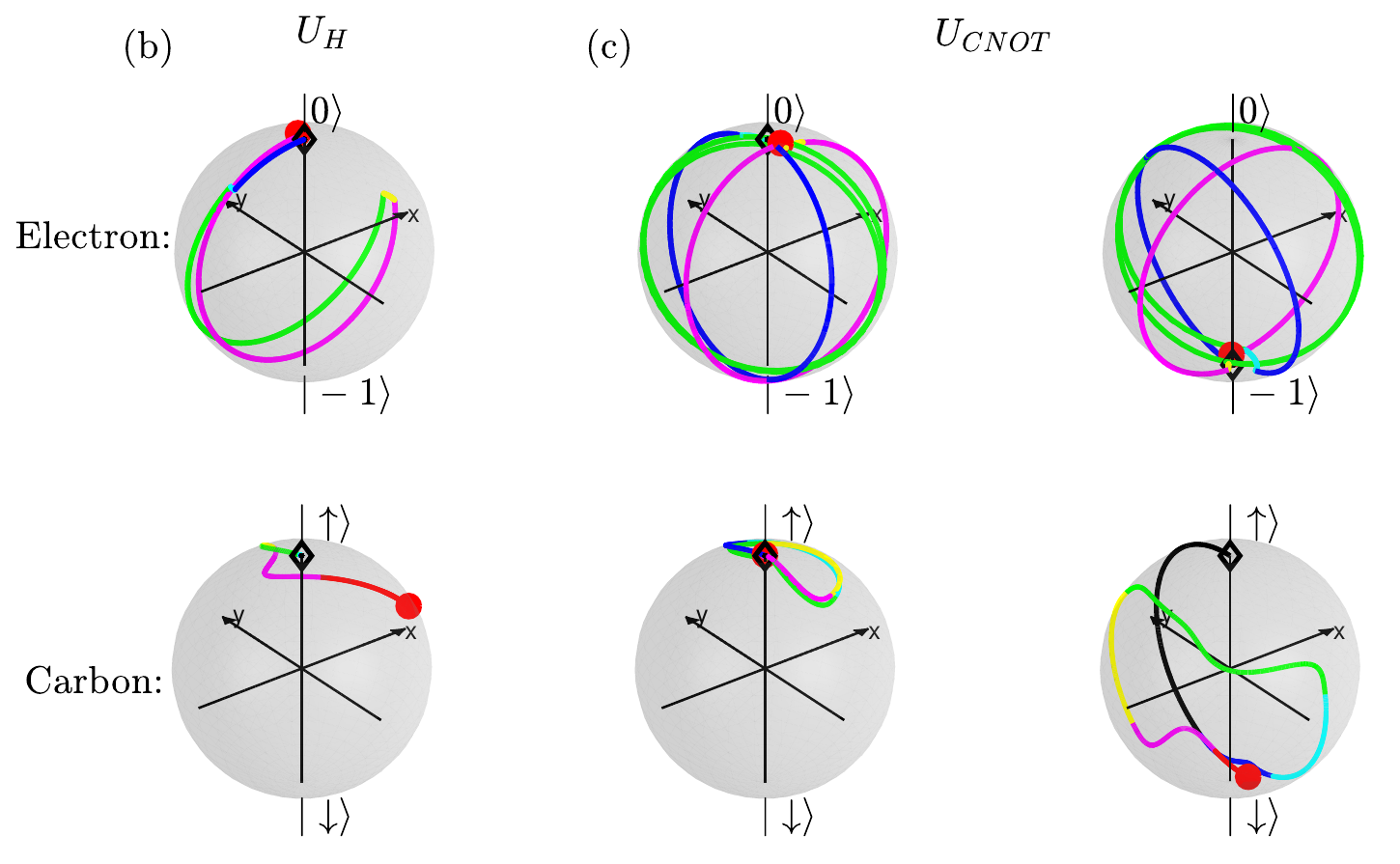}
\caption{(a) MW pulse sequence to realize $U_{H}$ and $U_{CNOT}$ by IC, at
a fixed $\omega_{1}$. The delays $\tau_{i}$, MW pulse durations
$t_{i}$ and phases $\phi_{i}$ are the free variables to be optimized.
(b, c) Evolution trajectories of electron and $^{13}$C upon the application
of $U_{H}$ and $U_{CNOT}$ for specific initial states. The diamond
indicates the initial state and the circle the final state.}
\label{bloch}
\end{figure}

\section*{2. Analytical form of pulse sequence to map the state $|0\uparrow\rangle$
to $|0\frac{(\uparrow+\downarrow)}{\sqrt{2}}\rangle$}

Here, we design an analytical form for the pulse sequence to map the
electron-$^{13}$C spin state from an initial state $|0\uparrow\rangle$
to a final state $|0\frac{(\uparrow+\downarrow)}{\sqrt{2}}\rangle$.
We choose a generic pulse sequence $(180^{\circ}-\tau_{1}-180^{\circ}-\tau_{2})$,
where the $180^{\circ}$ pulse acts on the electron that is resonant
with the ESR transition $0\leftrightarrow-1$, and $\tau_{i}$ are
delays. The unitary operator for the $180^{\circ}$ pulse is 
\[
U_{\pi}=e^{-i\pi I_{x}}.
\]

During the delays, $\tau_{i}$ with $i=1,2$, the system evolves under
the free evolution Hamiltonian ${\cal {H}}_{s}$, where 
\begin{equation}
{\cal H}_{s}/(2\pi)=(-\nu_{C}-{A_{zz}}/{2})(E_{2}\otimes I_{z})+A_{zz}(I_{z}^{e}\otimes I_{z})+A_{zx}(I_{z}^{e}\otimes I_{x})-{A_{zx}}/{2}(E_{2}\otimes I_{x}).\label{hs}
\end{equation}
Here $\nu_{c}=0.158$ MHz is the $^{13}$C spin Larmor frequency,
and $A_{zz}=-0.152$ MHz, $A_{zx}=0.11$ MHz are the hyperfine couplings
with the $^{13}$C spin.

The corresponding evolution operator in the basis $\{|0\uparrow\rangle,|0\downarrow\rangle,|-1\uparrow\rangle,|-1\downarrow\rangle\}$
during $\tau_{i}$ is

\[
U_{\tau_{i}}=\begin{bmatrix}e^{i\pi\nu_{C}\tau_{i}} & 0 & 0 & 0\\
0 & e^{-i\pi\nu_{C}\tau_{i}} & 0 & 0\\
0 & 0 & \cos(\pi\nu_{-}\tau_{i})+i\cos(\kappa_{-})\sin(\pi\nu_{-}\tau_{i}) & i\sin(\kappa_{-})\sin(\pi\nu_{-}\tau_{i})\\
0 & 0 & i\sin(\kappa_{-})\sin(\pi\nu_{-}\tau_{i}) & \cos(\pi\nu_{-}\tau_{i})-i\cos(\kappa_{-})\sin(\pi\nu_{-}\tau_{i})
\end{bmatrix},
\]
where $\nu_{-}=\sqrt{A_{zx}^{2}+(\nu_{C}+A_{zz})^{2}}$ is the $^{13}$C
spin transition frequency in the $m_{S}=-1$ subspace, and $\kappa_{-}=\tan^{-1}[A_{zx}/(A_{zz}+\nu_{C})]\approx86^{\circ}$
is the angle between the quantization axis of the $^{13}$C nuclear
spin and the NV axis.

The total propagator for the pulse sequence $(180^{\circ}-\tau_{1}-180^{\circ}-\tau_{2})$
is 
\[
U=U_{\tau_{2}}U_{\pi}U_{\tau_{1}}U_{\pi}.
\]
The state transformation $|0\uparrow\rangle$ $\rightarrow$ $|0\frac{(\uparrow+\downarrow)}{\sqrt{2}}\rangle$,
can be written as 
\[
|0(\uparrow+\downarrow)\rangle\langle0(\uparrow+\downarrow)|/2=U|0\uparrow\rangle\langle0\uparrow|U^{\dagger}.
\]

By equating the matrix elements $\langle0\uparrow|U|0\uparrow\rangle$
and $\langle0\uparrow|U|0\downarrow\rangle$ to $0.5$, we solve for
$\tau_{i}$: 
\begin{eqnarray}
\tau_{1} & = & \frac{1}{\pi\nu_{-}}\sin^{-1}(\frac{1}{\sqrt{2}\sin(\kappa_{-})})\\
\tau_{2} & = & \frac{1}{2\pi\nu_{C}}\cos^{-1}(\frac{\cos(\kappa_{-})}{\sin(\kappa_{-})})
\end{eqnarray}

where for our system, $\kappa_{-}\approx86^{\circ}$, $\nu_{-}=0.11$
MHz and thus $\tau_{1}=2.28$ $\mu$s, $\tau_{2}=1.53$ $\mu$s. $\tau_{1}+\tau_{2}$
sets the lower bound on the pulse sequence duration.

\section*{3. State evolution during to the pulse sequence to demonstrate Hadamard
gate}

We show the details of the state evolution during the pulse sequence
in Fig. 2(a) of the main manuscript. The $^{13}$C spin is initialized
into the state 
\begin{equation}
\rho_{0}^{c}=\frac{E_{2}}{2}+I_{z}.
\end{equation}
$E_{2}$ is a $2\times2$ identity matrix that does not evolve under
any operation, and we track the evolution of $I_{z}$ spin operator
at each stage of the pulse sequence \cite{cavanagh1995protein}. The
first Hadamard gate $U_{H}$ transforms $I_{z}$ as: 
\begin{equation}
I_{z}\longrightarrow I_{x}
\end{equation}

Since, initially the populations of the $m_{S}=-1$ subspace are zero,
we concentrate on the evolution of the $^{13}$C spin state in the
$m_{S}=0$ subspace, where the $^{13}$C spin Hamiltonian is 
\begin{equation}
{\cal H_{C}}=2\pi\nu_{c}I_{z}.
\end{equation}
Here $\nu_{c}$ is the $^{13}$C spin larmor frequency. During the
free precession for a duration $t$, $I_{x}$ evolves as 
\begin{equation}
I_{x}\longrightarrow I_{x}\cos(2\pi\nu_{c}t)+I_{y}\sin(2\pi\nu_{c}t)
\end{equation}
The second $U_{H}$ takes the above state to $I_{z}\cos(2\pi\nu_{c}t)-I_{y}\sin(2\pi\nu_{c}t).$
Thus the initial state $\rho_{0}^{c}$ goes to the final state 
\begin{equation}
\rho^{c}\longrightarrow\frac{E_{2}}{2}+I_{z}\cos(2\pi\nu_{c}t)-I_{y}\sin(2\pi\nu_{c}t)\label{eq:finalstate}
\end{equation}
The last clean-up operation transfers the population from $|0\downarrow\rangle$
to $|1\downarrow\rangle$. Hence, the remaining population of the
state $|0\uparrow\rangle$ of equation \eqref{eq:finalstate} is $[1+\cos(2\pi\nu_{c}t)]/2$.

\section*{4. State determination }

\begin{figure}[h]
\centering \includegraphics[width=7.5cm]{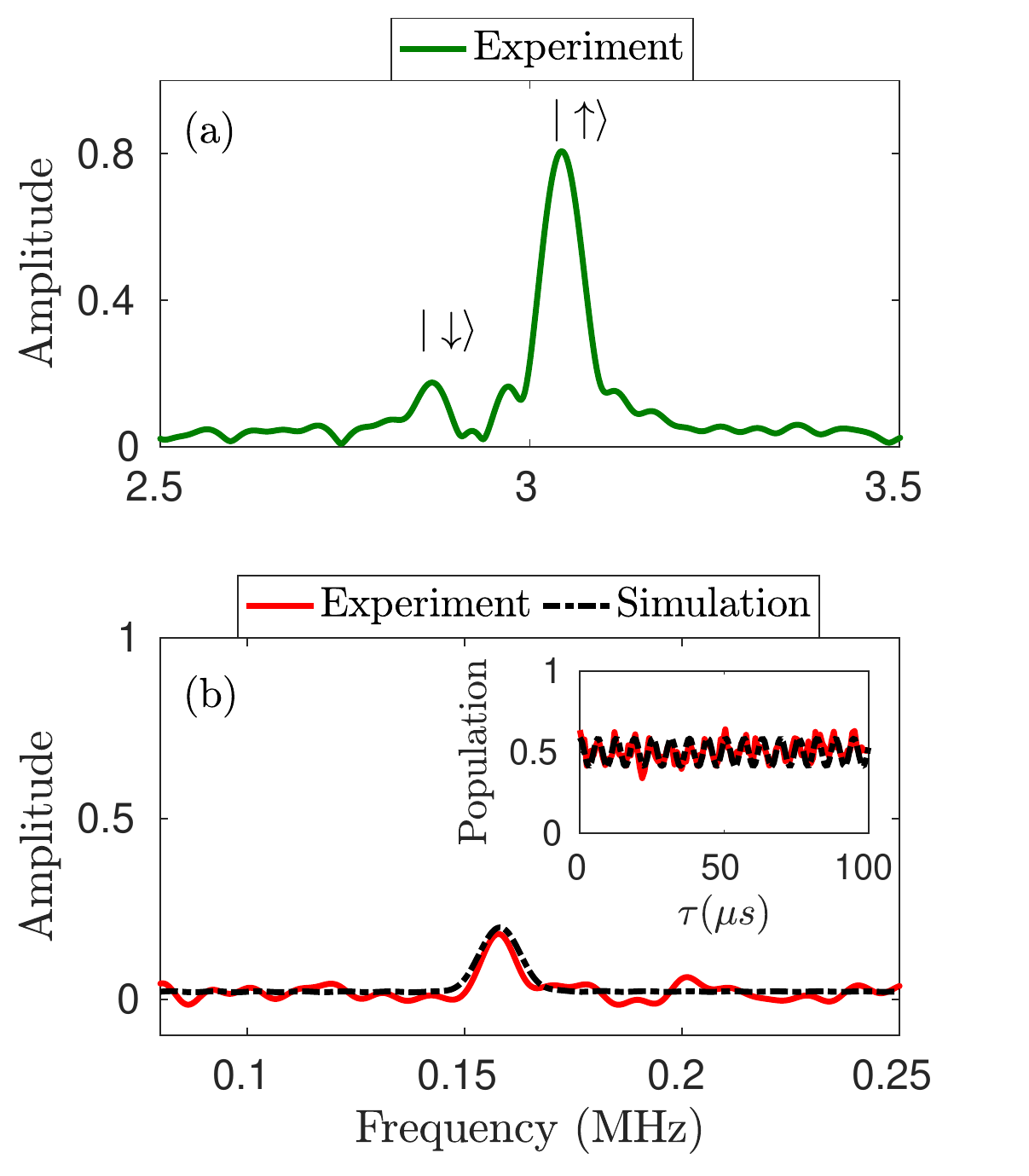}
\caption{(a) Population determination. ESR spectrum in the subspaces $m_{S}=\{0,1\}$
and $m_{N}=1$ showing $80\%$ population in state $|0\uparrow\rangle$
and $20\%$ population in state $|0\downarrow\rangle$. (b) Coherence
determination. The $^{13}$C spectrum corresponding to the pulse sequence
in Fig. 2(a) of the main manuscript without the first $U_{H}$ starting
from input state with $80\%$ population in state $|0\uparrow\rangle$
and $20\%$ population in state $|0\downarrow\rangle$. }
\label{state1} 
\end{figure}

Fig. \ref{state1}(a) shows the ESR spectrum of the state $\rho$.
It was obtained with the method described in \cite{zhang2018efficient}.
The two peaks correspond to the electron spin transitions in the $m_{S}=\{0,1\}$
subspace when the $^{13}$C spin is in the $|\uparrow\rangle$ and
$|\downarrow\rangle$ states, respectively, as indicated in the figure.
The area under these spectral lines is proportional to the populations
of $|0\uparrow\rangle$ and $|0\downarrow\rangle$. The analysis shows
that the populations of $|0\uparrow\rangle$ and $|0\downarrow\rangle$
are $\approx80\%$ and $\approx20\%$, respectively.

In order to calculate the coherence of the above state $\rho$, we
performed an experiment using the pulse sequence shown in Fig. 2(a)
of the main manuscript starting from a state with $\approx80\%$ and
$\approx20\%$ populations in states $|0\uparrow\rangle$ and $|0\downarrow\rangle$
respectively but omitted the first $U_{H}$ operation. If $\rho$
contains coherence between the states $|0\uparrow\rangle$ and $|0\downarrow\rangle$,
then these coherences will evolve during the free evolution time $t$.
The second $U_{H}$ converts one component of the coherence to population,
and the cleanup operation transfers population from state $|0\downarrow\rangle$
to $|1\downarrow\rangle$. Upon Fourier transformation of the remaining
populations of the state $|0\uparrow\rangle$ for variable $t$, we
get a frequency domain signal with a peak centered at the $^{13}$C
spin Larmor frequency $\nu_{c}$. Following this argument, we expect
no peak at $\nu_{c}$ in the absence of the above coherence terms.
The experimental result shown in Fig. \ref{state1}(b) indicated the
presence of a peak at $\nu_{c}$, thereby indicating the presence
of coherence between the states $|0\uparrow\rangle$ and $|0\downarrow\rangle$.
We determined this coherence by fitting the experimental population
of state $|0\uparrow\rangle$ as a function of variable delay $t$
(in Fig. 2(a) of the main manuscript) with the corresponding theoretical
input state and by optimizing the coherence amplitudes. We found a
coherence of 0.08 and thus our state before the clean-up operation
was 
\[
\rho=\left[{\begin{array}{cccc}
0.8 & 0.08 & 0 & 0\\
0.08 & 0.2 & 0 & 0\\
0 & 0 & 0 & 0\\
0 & 0 & 0 & 0
\end{array}}\right].
\]

We further purified this state by a clean-up operation that transferred
the population from $|0\downarrow\rangle$ to $|1\downarrow\rangle$
and the coherence between the states $\{|0\uparrow\rangle,|0\downarrow\rangle\}$
to the states $\{|0\uparrow\rangle,|1\downarrow\rangle\}$ \cite{zhangcnot}.
This clean-up is a MW pulse sequence $(90_{x}-\tau_{c}-90_{y})$,
where $90_{x/y}$ are pulses with rotation angle $90^{\circ}$ about
the $x/y$-axis applied to the $m_{S}=0\leftrightarrow1$ transition
with 0.5 MHz Rabi frequency and $\tau_{c}=1/(2|A_{zz}|)$ is the delay
between them. After this clean-up, our system subspace spanned by
$m_{S}=\{0,-1\}$ and $m_{N}=1$ was in the pure state

\begin{equation}
\psi_{0}=|0\uparrow\rangle
\end{equation}

\section*{5. Error estimation for CNOT}

In this section, we estimate the experimental fidelity of the state
after $U_{CNOT}$ using the results from Fig. 5 of the main text.
Here Fig. 5(a) corresponds to the case when the electron is in state
$|0\rangle$ and thus according to the definition of our gate operation,
$U_{CNOT}$ is an identity operation on the $^{13}$C spin. In the
case where the electron is in state $|-1\rangle$, $U_{CNOT}$ flips
the $^{13}$C spin and the results are shown in Fig. 5(b). The theoretical
$P_{0\downarrow}$ has the functional form $P_{0\downarrow}(\theta)=[1-\cos\theta]/2$,
where the angle $\theta$ parametrises the electron spin input states
before applying $U_{CNOT}$. For all $\theta=[0,2\pi]$, we matched
the experimental $P_{0\downarrow}$ by multiplying the corresponding
theoretical populations by $0.7$ and $0.9$ for $m_{S}=-1$ and $m_{S}=0$
respectively. Thus we observed a $10\%$ signal loss when the electron
was in state $|0\rangle$ and a $30\%$ signal loss when electron
was in state $|-1\rangle$. The average of these errors is $20\%$
and hence the experimental fidelity of the state after $U_{CNOT}$,
which in this case is calculated by measuring $P_{0\downarrow}$,
amounted to about $80\%$, in agreement with the results in Fig. 4
of the main manuscript where data are shown for $\theta$ = $\pi$.

\section*{6. Gates in Multiqubit systems}

Addressing and controlling individual qubits in multiqubit systems
is necessary to realize scalable quantum systems. The central electron
spin in the NV centers of diamond has potential to be coupled to multiple
$^{13}$C spins, thereby offering a possibility of realizing multiqubit
registers. However, the presence of these multiple nuclear spins also
is a main contribution to the decoherence and limits the spectral
resolution. The duration of the gate operations should therefore not
exceed the electron spin coherence time. We here extend our indirect
control scheme to the implementation of simple gate operations in
multiqubit systems consisting of up to 6 qubits, and check the typical
gate durations, the minimum required electron spin coherence time
and the control overhead.

Our $n$-qubit system consists of 1 electron spin, 1 $^{14}$N spin
and $(n-2)$ $^{13}$C spins. Here, the operations that we chose are
controlled-controlled rotations where the electron spin and the $^{14}$N
spin are the control qubits and an individual $^{13}$C spin is the
target qubit. On the remaining spins, the operation should implement
a unit operation (NOOP). In the rotating frame of the electron spin
with frequency given by $D+\nu_{e}-A_{N}$ (where the notations are
defined in the main text), the $n$-qubit system Hamiltonian in the
subspaces $m_{S}=\{0,-1\}$ and $m_{N}=1$ can be written as 
\begin{equation}
\frac{{\cal H}_{s}^{\otimes n}}{2\pi}=|1_{N}\rangle\langle1_{N}|\otimes(|0_{e}\rangle\langle0_{e}|\otimes\sum_{j=1}^{n-2}{\cal H}_{0}^{j}+|-1_{e}\rangle\langle-1_{e}|\otimes\sum_{j=1}^{n-2}{\cal H}_{-1}^{j}),
\end{equation}
where $e$ represents the electron, $N$ represents the $^{14}$N,
${\cal H}_{0}^{j}=-\nu_{C}I_{z}^{j}$ and ${\cal H}_{-1}^{j}=-(\nu_{C}+A_{zz}^{j})I_{z}^{j}-A_{zx}^{j}I_{x}^{j}$
are the Hamiltonians for the $j^{\mathrm{th}}$ $^{13}$C spin. The
$^{13}$C spin Larmor frequency is $\nu_{C}=0.158$ MHz and the chosen
hyperfine couplings with the $^{13}$C spins are listed in Table \ref{tab3}.
The simulated spectrum of this Hamiltonian for $n=6$ in the $m_{S}=\{0,-1\}$
and $m_{N}=1$ subspaces is shown in Fig. \ref{fid}. Thus we see
that, a minimum $T_{2}{}^{*}=1/(\pi\delta\nu)\approx30\,\mu$s, where
$\delta\nu$ is the line width of the ESR spectra, is necessary to
spectrally address individual $^{13}$C spins in this system.

\begin{table}[th]
\centering %
\begin{tabular}{|c|c|c|}
\hline 
$j$  & $A_{zz}^{j}$ (MHz)  & $A_{zx}^{j}$ (MHz) \tabularnewline
\hline 
1  & $A_{zz}^{1}=-0.152$  & $A_{zx}^{1}=0.110$ \tabularnewline
\hline 
2  & $A_{zz}^{2}=(1.5)\cdot A_{zz}^{1}$  & $A_{zx}^{2}=(1.5)\cdot A_{zx}^{1}$ \tabularnewline
\hline 
3  & $A_{zz}^{3}=(2/3)\cdot A_{zz}^{1}$  & $A_{zx}^{3}=(2/3)\cdot A_{zx}^{1}$ \tabularnewline
\hline 
4  & $A_{zz}^{4}=(2.5)\cdot A_{zz}^{1}$  & $A_{zx}^{4}=(2.5)\cdot A_{zx}^{1}$ \tabularnewline
\hline 
\end{tabular}\caption{Hyperfine couplings $A_{zz}^{j},A_{zx}^{j}$ for a system of four
$^{13}$C spins. $j=1\dots4$ represents the label for the $^{13}$C
spins.}
\label{tab3} 
\end{table}

\begin{figure}[h]
\centering \includegraphics[width=10.5cm]{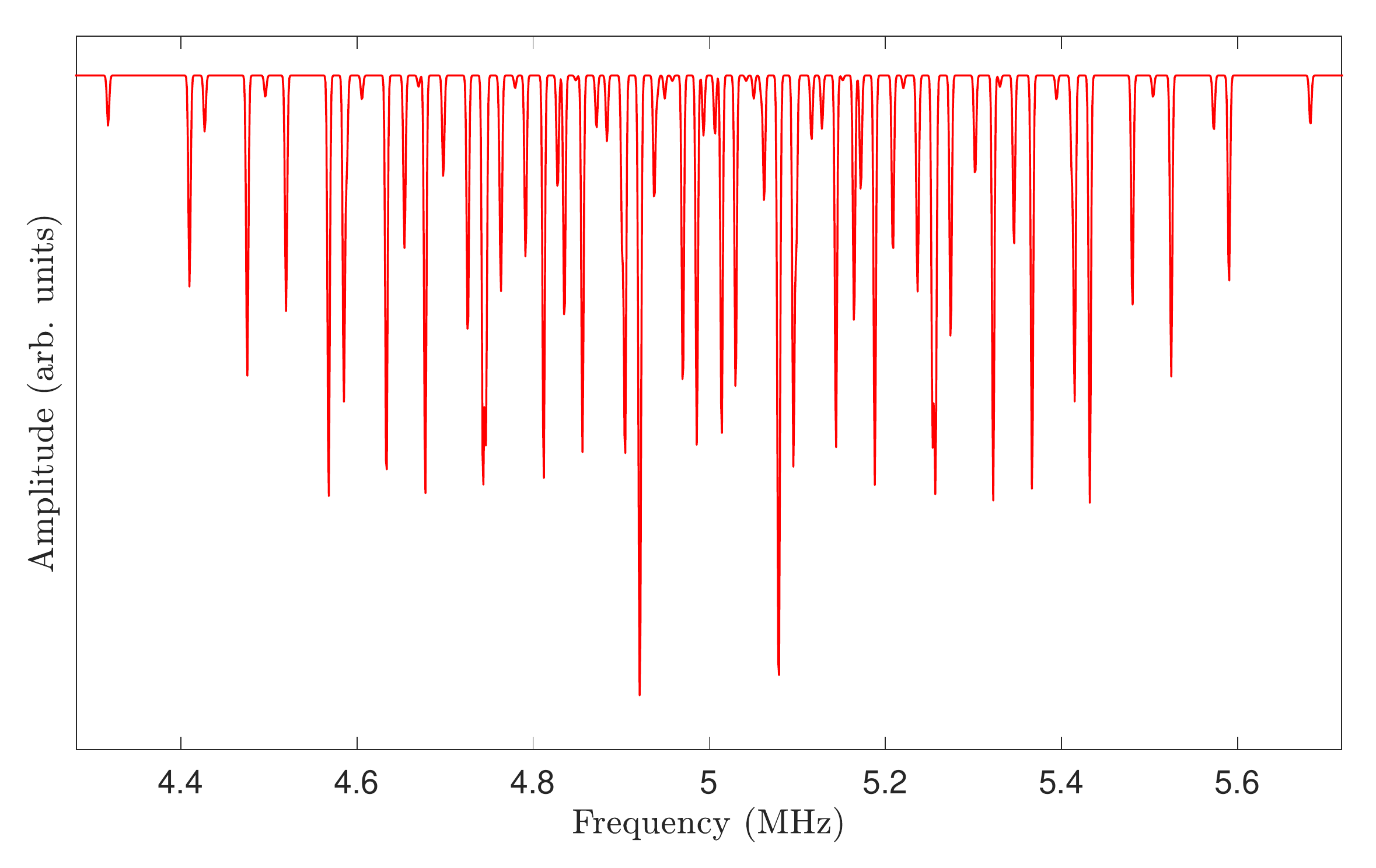}
\caption{Simulated ESR spectrum of the $n=6$ system, i.e, with all the four
$^{13}$C spins, in the $m_{S}=\{0,-1\}$ and $m_{N}=1$ subspace.
The coupling parameters are listed in Table \ref{tab3} and the detuning
frequency was set to 5 MHz.}
\label{fid} 
\end{figure}

To design pulse sequences for arbitrary gate operations in multiqubit
systems, we extend the optimization protocol for a two qubit system
as explained in the main text to that of an $n$ qubit system with
the system Hamiltonian ${\cal H}_{s}^{\otimes n}$. The control (MW)
Hamiltonian in $m_{N}=1$ is $\omega_{1}[\cos\phi_{i}(s_{x}\otimes E^{\otimes{2^{n-2}}})+\sin\phi_{i}(s_{y}\otimes E^{\otimes{2^{n-2}}}]$
where $\omega_{1}$ is the MW pulse amplitude and $E^{\otimes{m}}$
is the ${2^{m}}\times{2^{m}}$ identity matrix. We show that the controlled-controlled
rotations can be implemented using the generic 4 pulse MW pulse sequence
as shown in Fig. \ref{pulse}. As explained in the main text, $(\tau_{i},t_{i},\phi_{i})$
are the pulse sequence parameters that are to be optimized to design
gates with maximum fidelity with a target unitary operator. The Rabi
frequency $\omega_{1}/(2\pi)$ is set to $0.5$ MHz which is used
to select the $m_{N}=1$ subspace of the $^{14}$N spin and the pulses
are not selective to any of the $^{13}$C spin transitions.

\begin{figure}[h]
\centering \includegraphics[width=10.5cm]{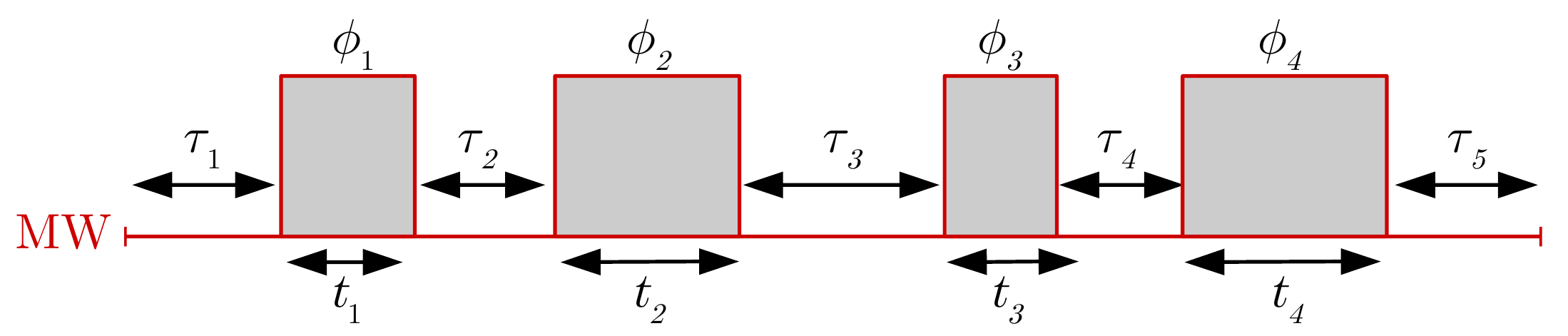}
\caption{MW pulse sequence to implement controlled-controlled rotations on
individual $^{13}$C spins in muti-qubit systems up to at least $n=6$. }
\label{pulse} 
\end{figure}

We first simulate a controlled-controlled NOT gate in a $n=4$ system.
We separately implement two controlled-controlled NOT gates targeting
the $j=1$ spin on two four-qubit systems with different $^{13}$C
spin hyperfine couplings as indicated in Figs. \ref{circuits_lab}(a,
b). In Fig. \ref{circuits_lab}(a), the system consists of 1 e, 1
$^{14}$N, and $j=1,2$ carbon spins where the hyperfine coupling
with the $j=2$ spin is larger than that of the $j=1$ spin. In Fig.
\ref{circuits_lab}(b), we choose a system with 1 e, 1 $^{14}$N,
and $j=1,3$ carbon spins where the hyperfine coupling with the $j=3$
spin is weaker than that of the $j=1$ spin. The optimized pulse sequence
parameters corresponding to the sequence in Fig. \ref{pulse} are
shown in Table \ref{tab2}. The MW pulse sequence for implementing
the controlled-controlled-NOT gate targeting the $j=1$ carbon spin
and NOOP on the other carbon spin in either of the two cases were
efficiently designed using only 4 MW pulses with total duration of
the sequence less than $15\,\mu$s and the theoretical gate fidelities
were greater than $0.99$.

\begin{figure}[h]
\centering \includegraphics[width=10cm]{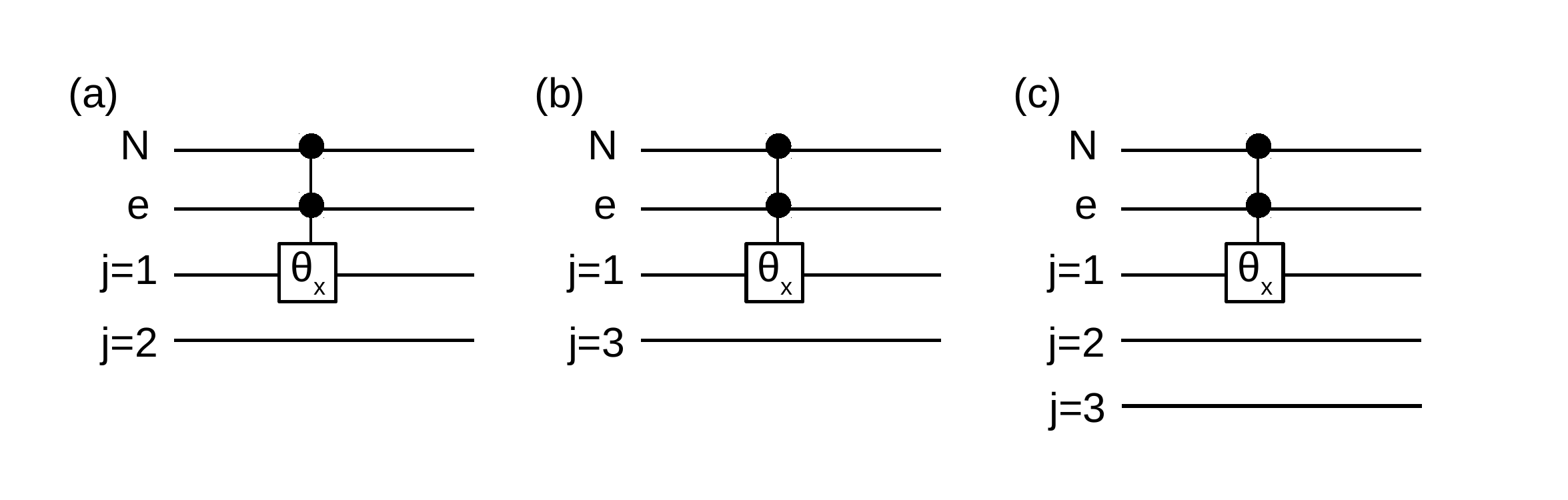}
\caption{Circuit to implement controlled-controlled-rotations: (a, b) in $n=4$
and (c) in $n=5$. Here $\theta_{x}=e^{-i\theta I_{x}}$.}
\label{circuits_lab} 
\end{figure}

The $n=5$ system consists of 1 e, 1 $^{14}$N, and $j=1,2,3$ carbon
spins. Fig. \ref{circuits_lab}(c) shows the circuit for implementing
a selective controlled-controlled-NOT gate targeting only the $j=1$
spin. The corresponding 4-pulse MW pulse sequence parameters are shown
in Table \ref{tab2}. This MW pulse sequence implements the above
controlled-controlled-NOT gate with a fidelity greater than $0.98$
within a duration of $15\,\mu$s, while simultaneously implementing
NOOP on the $j=2,3$ spins.

\begin{figure}[b]
\centering \includegraphics[width=12cm]{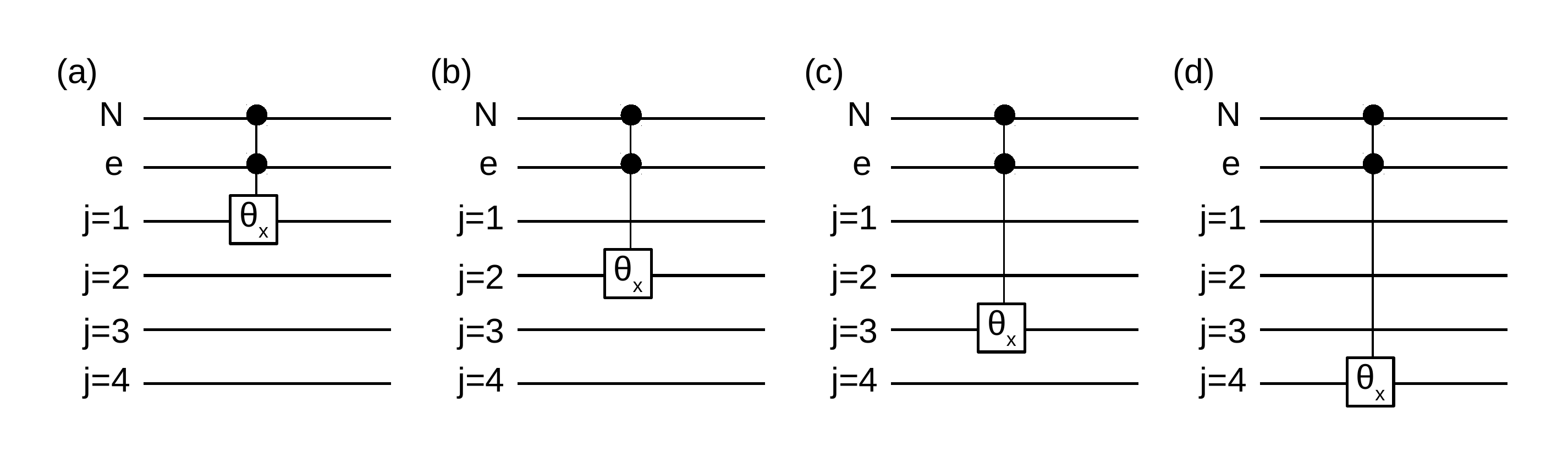}
\caption{Circuits to implement controlled-controlled-rotations in systems with
$n=6$ qubits. Here $\theta_{x}=e^{-i\theta I_{x}}$.}
\label{circuits_lab1} 
\end{figure}

Finally, we consider the $n=6$ qubit case with 1 e, 1 $^{14}$N,
and $j=1,2,3,4$ carbon spins. Here we optimize the parameters of
the pulse sequence in Fig. \ref{pulse} for 4 controlled-controlled-rotation
operations using the different $^{13}$C spins as target qubits, as
shown in Fig. \ref{circuits_lab1}. Table \ref{tab2} gives the resulting
pulse sequence parameters for each of these cases with theoretical
gate fidelities ranging from $0.93$ to $0.99$ and total durations
ranging from 22-28 $\mu$s. As an example, the form of the ideal and
simulated operator in the $m_{N}=1$ subspace corresponding to Fig.
\ref{circuits_lab1}(a) is shown in Fig. \ref{U6q}.

\begin{table}[h]
\centering %
\begin{tabular}{|c|c|c|c|c|c|c|c|c|c|c|c|c|c|c|c|c|c|c|}
\hline 
$n$  & $j$  & Figure  & $\theta$  & $\tau_{1}$  & $\tau_{2}$  & $\tau_{3}$  & $\tau_{4}$  & $\tau_{5}$  & $t_{1}$  & $t_{2}$  & $t_{3}$  & $t_{4}$  & $\phi_{1}$  & $\phi_{2}$  & $\phi_{3}$  & $\phi_{4}$  & Fidelity  & Gate duration ($\mu$s)\tabularnewline
\hline 
$4$  & 1,2  & \ref{circuits_lab}(a)  & $180^{\circ}$  & $1.05$  & $1.04$  & $1.09$  & $1.07$  & $4.13$  & $2.44$  & $1.12$  & $0.94$  & $1.49$  & $61^{\circ}$  & $270^{\circ}$  & $233^{\circ}$  & $90^{\circ}$  & $0.991$  & $14.4$\tabularnewline
\hline 
$4$  & 1,3  & \ref{circuits_lab}(b)  & $180^{\circ}$  & $0.43$  & $0.83$  & $0.79$  & $1.08$  & $2.55$  & $2.14$  & $1.36$  & $1.80$  & $1.46$  & $298^{\circ}$  & $218^{\circ}$  & $252^{\circ}$  & $90^{\circ}$  & $0.996$  & $12.4$\tabularnewline
\hline 
$5$  & 1,2,3  & \ref{circuits_lab}(c)  & $180^{\circ}$  & $2.35$  & $2.13$  & $3.99$  & $0.63$  & $0.48$  & $1.51$  & $1.93$  & $0.25$  & $1.27$  & $296^{\circ}$  & $315^{\circ}$  & $181^{\circ}$  & $90^{\circ}$  & $0.983$  & $14.5$\tabularnewline
\hline 
$6$  & 1,2,3,4  & \ref{circuits_lab1}(a)  & $180^{\circ}$  & $4.27$  & $2.22$  & $0.79$  & $3.91$  & $6.14$  & $1.34$  & $1.01$  & $1.65$  & $1.16$  & $206^{\circ}$  & $129^{\circ}$  & $325^{\circ}$  & $90^{\circ}$  & 0.989  & 22.5\tabularnewline
\hline 
$6$  & 1,2,3,4  & \ref{circuits_lab1}(b)  & $180^{\circ}$  & $4.18$  & $7.26$  & $1.83$  & $1.06$  & $6.38$  & $2.35$  & $0.95$  & $0.59$  & $0.24$  & $182^{\circ}$  & $170^{\circ}$  & $245^{\circ}$  & $90^{\circ}$  & 0.939  & 24.8\tabularnewline
\hline 
$6$  & 1,2,3,4  & \ref{circuits_lab1}(c)  & $45^{\circ}$  & $5.01$  & $2.02$  & $2.07$  & $3.72$  & $5.17$  & $0.50$  & $1.89$  & $0.95$  & $0.93$  & $276^{\circ}$  & $262^{\circ}$  & $254^{\circ}$  & $90^{\circ}$  & 0.970  & 22.3\tabularnewline
\hline 
$6$  & 1,2,3,4  & \ref{circuits_lab1}(d)  & $45^{\circ}$  & $4.83$  & $3.77$  & $4.45$  & $2.58$  & $6.75$  & $1.68$  & $1.98$  & $1.54$  & $0.33$  & $176^{\circ}$  & $76^{\circ}$  & $97^{\circ}$  & $90^{\circ}$  & 0.976  & 27.9\tabularnewline
\hline 
\end{tabular}\caption{Optimized pulse sequence parameters $(\tau_{i},t_{i},\phi_{i})$ corresponding
to the pulse sequence in Fig. \ref{pulse} to implement controlled-controlled
rotations on individual $^{13}$C spins in the system of size $n$.
$j$ indicates the different $^{13}$C spins that are considered in
each case. Each row corresponds to a specific operation as indicated
by the Fig. \ref{circuits_lab} or \ref{circuits_lab1}. In all these
cases, $\omega_{1}/2\pi=0.5$ MHz. Fidelity represents the theoretically
calculated gate fidelities.}
\label{tab2} 
\end{table}

\textit{Efficiency and comparison with methods based on DD cycles.--}
The numerically optimized pulse sequence parameters for implemening
controlled-controlled rotation gates between specific pairs of qubits
in systems with up to $n=6$ qubits show that the indirect control
scheme proposed in this work is efficient with only 4 MW pulses and
with theoretical gate fidelities ranging from $0.93$ to 0.99. As
can be seen in Table \ref{tab2}, the gate durations gradually increase
from about $12\,\mu$s for $n=4$ up to $28\,\mu$s for $n=6$. These
gate durations will further increase (about $2-3\,\mu$s) if the sequences
are made robust with respect to the deviations in $\omega_{1}$. Thus
as seen in Table \ref{tab2}, a mimimum $T_{2}^{*}$ of about $30\,\mu$s
is necessary to implement controlled-controlled rotations in the $n=6$
system that we considered. Also the control overhead was only 4 MW
pulses.

The $^{12}$C enriched NV sample that we used in our experiments had
an electron spin $T_{2}^{*}$ of about $20\,\mu$s and the electron
spin coherence time $T_{2}$ for this sample was more than $1.3$$\,$ms
\cite{zhang2018bloch}. This $T_{2}^{*}$ is sufficient for indirect
control a single $^{13}$C spin, but $T_{2}^{*}$ of the electron
spin is shorter in crystals with higher $^{13}$C spin concentration
\cite{mizuochi2009coherence} and the total gate durations will exceed
the $T_{2}^{*}$. In such cases, protected quantum gates that are
interleaved with the DD pulses \cite{zhang2014protected} so as to
extend the electron spin $T_{2}^{*}$ beyond $30-100\,\mu$s would
assist in coherently addressing the individual nuclear spins in larger
spin systems. Also, in our previous work, we showed that one can further improve the fidelity and gate duration by polarizing the $^{14}$N spin instead of working in the subspace  $m_N=1$ \cite{zhang2018efficient}. For $n>10$, optimizing the pulse sequence parameters
using classical computers gets increasingly difficult. Nevertheless,
our control scheme could be very useful in cases like Ref. \cite{jiang2007distributed}
where it has been shown that only 5 qubits are sufficient to realize
a fully functional quantum repeater node.

Our scheme is efficient for the systems where $\nu_{C}$ is comparable
to the hyperfine couplings. In such cases, the difference $\delta$
between the orientation of the quantization axes of the $^{13}$C
spin with the NV axis in $m_{S}=0$ and 1 subspaces is close to $\pi/2$.
Following this, the low control overhead of only 4 MW pulses derives
from the argument that any rotation in the SO(3) group can be constructed
with $\le m+2$ rotations where $\pi/(m+1)\le\delta<\pi/m$ \cite{khaneja2007switched,lowenthal1971uniform}.
Our scheme holds even for systems where the hyperfine couplings are
only a few tens of kHz. This requires that the multiple $^{13}$C
spins under consideration have similar coupling strengths. One can
then adjust the external static magnetic field to bring $\nu_{C}$
to a value that is comparable with the couplings. Thus, using this
scheme, even very weakly coupled $^{13}$C spins could be controlled
with as few as 3 MW pulses. On the other hand, the indirect control
methods based on multiple cycles of DD sequences to achieve a desired
nuclear spin rotation work in a different regime where $\nu_{C}\gg(A_{zz},A_{zx})$
\cite{taminiau2012detection,taminiau2014universal}. The latter method
requires tens to hundreds of MW pulses.

\begin{figure}[bh]
\centering \includegraphics[width=8cm]{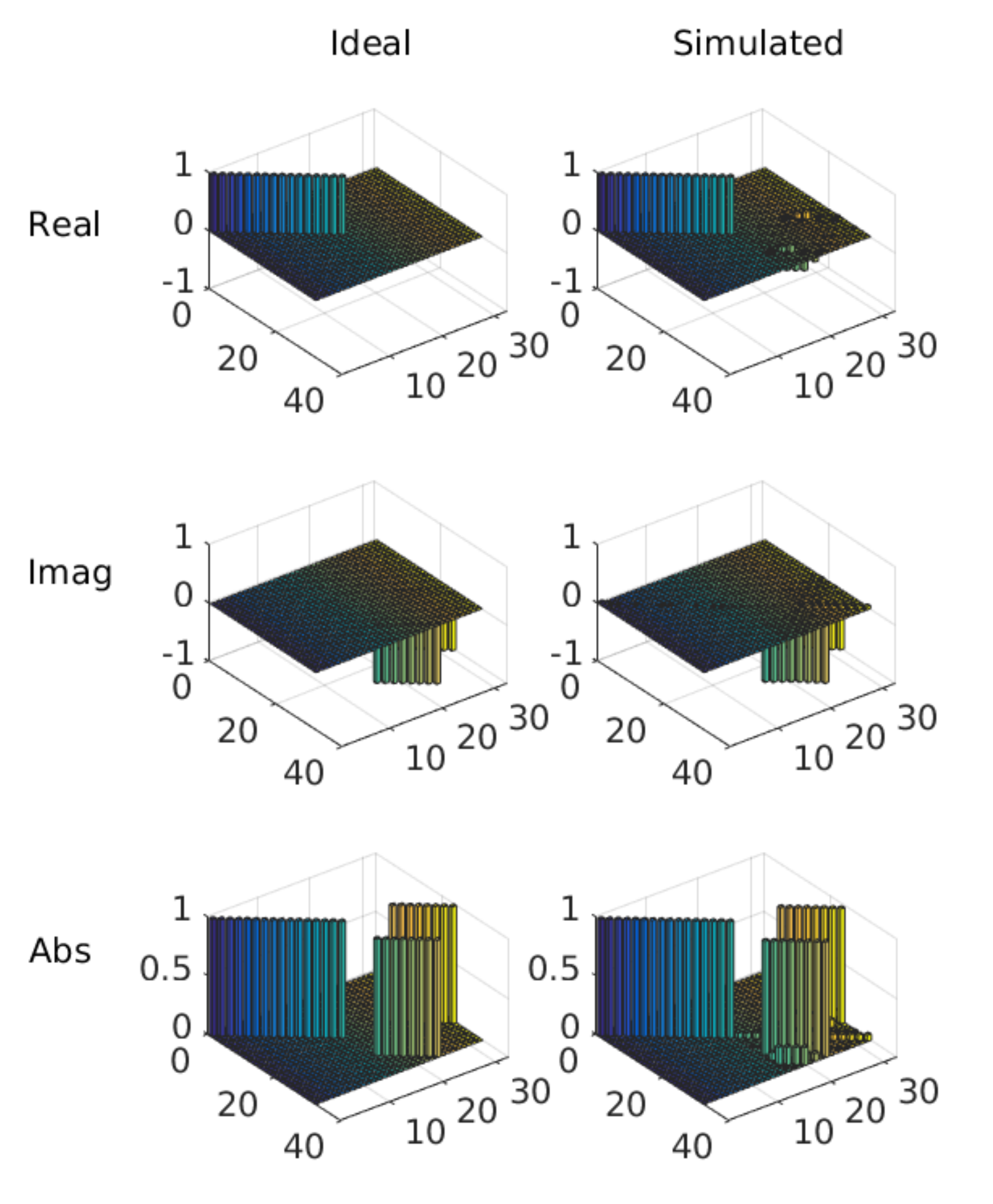}
\caption{Ideal (right) and simulated (left) unitaries in the $m_{N}=1$ subspace
with 1 e and four $^{13}$C spins using Fig. \ref{circuits_lab1}(a).
The pulse sequence parameters for the simulated unitary are given
in Table \ref{tab2}. In the simulation, $\omega_{1}$ was set to
0.5 MHz and hence, for simplicity, we here show the reduced unitary
in the $m_{N}=1$ subspace.}
\label{U6q} 
\end{figure}

\section*{7. Spatial distance between the electron and the $^{13}$C spin.}

The dipolar Hamiltonian between the electron spin with $\gamma_{e}=-1.761\times10^{11}$
rad $s^{-1}T^{-1}$ and a $^{13}$C spin with $\gamma_{C}=6.728\times10^{11}$
rad $s^{-1}T^{-1}$ which is located at a distance $r$ from electron
is 
\begin{equation}
\mathcal{H}_{d}=\vec{S}\cdot\bar{A}\cdot\vec{I}=-\frac{\mu_{0}}{4\pi}\frac{\gamma_{e}\gamma_{C}h}{r^{3}}[3(\vec{S}.\hat{n})(\vec{I}.\hat{n})-\vec{S}\cdot\vec{I}]\label{Hdip}
\end{equation}
Here $\bar{A}$ if the hyperfine tensor, $\vec{S}$ and $\vec{I}$
are the electron and $^{13}$C spin operators respectively, $h=6.626\times10^{-34}\,$Js
is Planck's constant, $\mu_{0}=4\pi\times10^{-7}\,$Hm$^{-1}$ is
the magnetic permeability, and $\hat{n}=[n_{x},n_{y},n_{z}]=[\sin(\theta)\cos(\phi),\sin(\theta)\sin(\phi),\cos(\theta)]$
is a unit vector pointing from the electron to the$^{13}$C. 

By equating the coefficients of $S_{z}I_{z}$ and $S_{z}I_{x}$ in
Eq. \ref{Hdip}, we get 
\begin{equation}
\begin{split} & A_{zz}=-0.152\,\mathrm{MHz}=[b(r)/2\pi].[3\cos^{2}(\theta)-1].\\
 & A_{zx}=0.110\,\mathrm{MHz}=[b(r)/2\pi].[3\sin(\theta)\cos(\theta)]
\end{split}
\end{equation}
where $b(r)=-\frac{\mu_{0}}{4\pi}\frac{\gamma_{e}\gamma_{C}h}{r^{3}}$
and we have set $\phi=0$ by chosing a reference frame in which the
$^{13}$C is located in the zx-plane. By solving the above equations,
we detemined the spatial distance between the electron and $^{13}$C
spin as $r=0.8924$ nm and $\theta=78^{\circ}$.

\section*{8. Effects of operations in the $^{14}$N subspaces $m_{N}=\{0,1,-1\}$}

In this section, we show that our operations on the electron spin
in the subpaces $m_{S}=\{0,-1\}$ and $m_{N}=1$ to implement rotations
on the $^{13}$C spin do not effect the other $^{14}$N subpaces $m_{N}=\{0,-1\}$.
To demonstrate this, we compare the thermal state ESR spectrum with
the pure state ESR spectrum. As explained in the main text, the pure
state is obtained by initializing the electron to state $|0\rangle$
by a 532 nm laser pulse and the$^{13}$C spin is initialized to $|0\rangle$
by the indirect control method. The corresponding MW pulse sequence
driving the electron spin consists of 3 pulses followed by a laser
pulse of duration 1.1 $\mu$s as explained in the main text. As with
the gate implementations for Hadamard and CNOT, the MW pulse amplitude
was set to 0.5 MHz and the optimized pulse sequence parameters were
$(\tau_{1},\tau_{2},\tau_{3},\tau_{4},t_{1},t_{2},t_{3},\phi_{1},\phi_{2},\phi_{3})=(0,2.09\,\mu s,2.59\,\mu s,0.84\,\mu s,0.52\,\mu s,0.45\,\mu s,1.03\,\mu s,16^{\circ},108^{\circ},90^{\circ})$.

Fig. \ref{n} shows the experimental ESR spectrum for the thermal
state (top trace) and pure state (bottom trace) in the $m_{S}=\{0,-1\}$
subspace. The $^{14}$N spin subspaces $m_{N}$ are indicated. In
each $m_{N}$ subspace, the thermal spectra have four ESR transitions
as explainined in section 1 of this supplementary material. The numbers
1, 2, 3, 4 in the $m_{N}=1$ subspace mark the ESR transitions $|0\uparrow\rangle\leftrightarrow|-1\psi_{-}\rangle$,
$|0\uparrow\rangle\leftrightarrow|-1\varphi_{-}\rangle$, $|0\downarrow\rangle\leftrightarrow|-1\psi_{-}\rangle$,
$|0\downarrow\rangle\leftrightarrow|-1\varphi_{-}\rangle$ respectively.
The pure state spectrum contains only the 2 ESR lines, 1 and 2, in
the $m_{N}=1$ subspace but with almost twice the amplitude as in
the thermal state, consistent with the subspaces where our gates were
designed.

We see that the electron and $^{13}$C spins were polarized only in
the $m_{N}=1$ subspace while the other subspaces. e.g. $m_{N}=0$
retain all the four peaks with comparable spectral amplitudes. This
shows that our MW pulse sequences do not affect the $^{13}$C spins
in the other $^{14}$N spin subspaces.

\begin{figure}[h]
\centering \includegraphics[width=14cm]{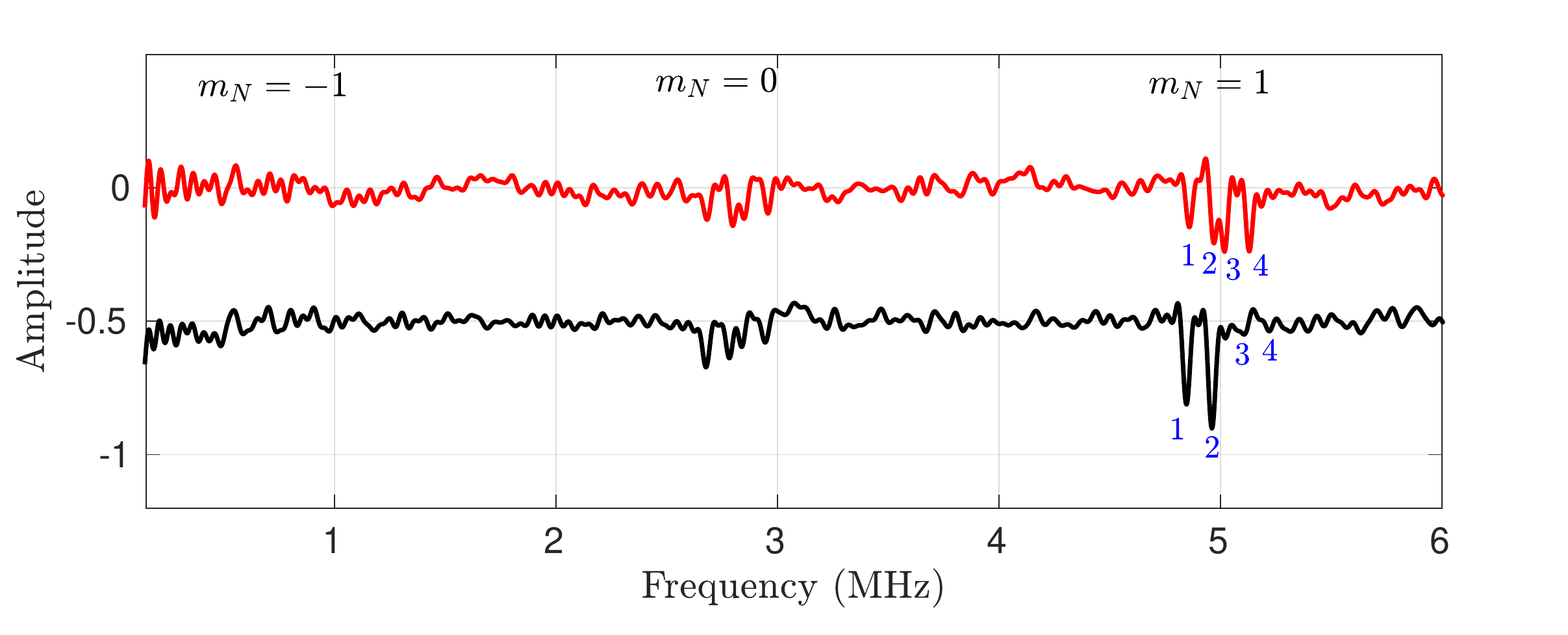}
\caption{ESR spectrum of the thermal state (top) and the pure state (bottom).
Here, the detunig frequency was set to 5 MHz. 1, 2, 3, 4 are the ESR
transitions in the $m_{N}=1$ subspace.}
\label{n} 
\end{figure}

\end{document}